%% file: main.tex
\documentclass[fleqn,usenatbib]{mnras}

\usepackage{newtxtext,newtxmath}
\usepackage[T1]{fontenc}
\usepackage{ae,aecompl}

\usepackage{graphicx}   % Including figure files

\usepackage{color,soul}  % Highlight changes following referee's report

\input{thesis_macros}

\title[Model quasar colours]{Modelling type 1 quasar colours in the era of Rubin and Euclid}

\author[Temple, Hewett \& Banerji]{Matthew J. Temple,$^{1,2}$\thanks{E-mail: Matthew.Temple@mail.udp.cl}
Paul C. Hewett$^{2}$
and Manda Banerji$^{3,2,4}$
\\
$^{1}$N\'ucleo de Astronom\'ia, Universidad Diego Portales, Av. Ej\'ercito Libertador 441, Santiago 8320000, Chile\\
$^{2}$Institute of Astronomy, University of Cambridge, Madingley Road, Cambridge CB3 0HA, UK\\
$^{3}$School of Physics \& Astronomy, University of Southampton, Southampton SO17 1BJ, UK\\
$^{4}$Kavli Institute for Cosmology, University of Cambridge, Madingley Road, Cambridge CB3 0HA, UK\\
}

\date{Accepted 2021 September 6. Received 2021 September 2; in original form 2021 July 8}

\pubyear{2021}

\begin{document}
\label{firstpage}
\pagerange{\pageref{firstpage}--\pageref{lastpage}}
\maketitle

%max length 250 words (200 for letters) Currently 235 words.
\begin{abstract}
%Abstract word limit of 250.
%The abstract should be presented as a single paragraph and briefly summarize the goals, methods, and new results presented in the paper.
We construct a parametric SED model which is able to reproduce the average observed SDSS-UKIDSS-\textit{WISE} quasar colours  to within  one tenth of a magnitude across a wide range of redshift ($0<z<5$) and luminosity ($-22>M_i>-29$).
This model is shown to provide accurate predictions for the colours of known quasars which are less luminous than those used to calibrate the model parameters, and also those at higher redshifts $z>5$.
Using a single parameter, the model encapsulates an up-to-date understanding of the intra-population variance in the rest-frame ultraviolet and optical emission lines of luminous quasars. At fixed redshift, there are systematic changes in the average quasar colours with apparent \textit{i}-band magnitude, which we find to be well explained by the contribution from the host galaxy and our parametrization of the emission-line properties. By including redshift as an additional free parameter, the model could be used to provide photometric redshifts for individual objects. For the population as a whole we find that the average emission line and host galaxy contributions can be well described by simple functions of luminosity which account for the observed changes in the average quasar colours across $18.1<i_\textrm{AB}<21.5$. 
We use these trends to provide predictions for quasar colours at the luminosities and redshifts which will be probed by the Rubin Observatory LSST and ESA-\textit{Euclid} wide survey.
The model code is applicable to a wide range of upcoming photometric and spectroscopic surveys, and is made publicly available.
\end{abstract}

% Select between one and six entries from the list of approved keywords.
% Don't make up new ones.

%%%%%%%%%%%%%%%%%%%%%%%%%%%%%%%%%%%%%%%%%%%%%%%%%%

%%%%%%%%%%%%%%%%% BODY OF PAPER %%%%%%%%%%%%%%%%%%
%This is a simple template for authors to write new MNRAS papers.
%See \texttt{mnras\_sample.tex} for a more complex example, and \texttt{mnras\_guide.tex}
%for a full user guide.

\section{Introduction}

\subsection{Wide-field extragalactic surveys in the 2020s}

Over the past few decades, larger and deeper surveys of the sky have each built a more complete census of the quasars and active galactic nuclei (AGN) which populate our universe, from the 2dF QSO Redshift Survey \citep{2000MNRAS.317.1014B, 2001MNRAS.322L..29C, 2004MNRAS.349.1397C}  through iterations of the Sloan Digital Sky Survey \citep[SDSS;][]{2000AJ....120.1579Y, 2002AJ....123.2945R, 2010AJ....139.2360S, 2017A&A...597A..79P, DR16Q}.

Upcoming photometric surveys such as the Vera C. Rubin Observatory Legacy Survey of Space and Time \citep[LSST;][]{2019ApJ...873..111I} and ESA's \textit{Euclid} mission \citep{2011arXiv1110.3193L, 2019A&A...631A..85E} will probe new regimes; both pushing fainter in luminosity and also providing  unprecedented amounts of data on variability in the time domain. 
In order to identify as many quasars as possible within these rich upcoming data-sets, we first need to understand what we expect new AGN to look like in those surveys. 
LSST and \textit{Euclid} will probe deeper than ever before and in so doing will detect tens of millions of new AGN. These apparently fainter AGN will be located in hitherto unpopulated regions of parameter space: both bright objects at higher redshifts and intrinsically fainter objects at high and low redshifts, whether that be due to lower mass black holes or lower accretion rates. For the lower luminosity AGN the contribution from starlight in the host galaxy to the total observed flux will be more significant.

At the same time, spectroscopic instruments such as VLT-MOONS  \citep{2020Msngr.180...24M} and 4MOST \citep{2019Msngr.175...42M} will conduct surveys of the extragalactic sky. Efficiently selecting AGN for these spectroscopic surveys is crucial in order to allow the community to address the most fundamental questions in studies of galaxy and supermassive black hole (SMBH) evolution, and to maximise the scientific return from the upcoming generation of wide-field photometric surveys.

\subsection{The need for parametric SED models}

The upcoming generation of wide-field surveys will probe a  large dynamic range in apparent magnitude, corresponding to a broader sub-domain of the AGN luminosity function (at any given redshift) than has been sampled to date. Even at fixed redshift the population will exhibit a range of observed quasar properties. Quantifying the intra-population properties with the minimum number of parameters will be a prerequisite for many statistical investigations. Current model spectral energy distributions (SEDs) for AGN based on a single fixed template or composite spectrum will not prove adequate.
Here we aim to model the SEDs of luminous, type 1 AGN (quasars) in a way that encapsulates the systematic changes in observed optical and near-infrared properties as a function of luminosity, as well as redshift.
Notably, we chose parameters which are motivated by known astrophysics, such as the systematic changes in the emission line properties and the relative contribution of the host galaxy. The model allows parameters to be specified independently to explore the intra-population variance at fixed redshift and luminosity. 

Since the structure of SMBH accretion discs is not fully understood in a quantitative sense, the majority of existing SED templates are purely empirical.
The empirical mean quasar template of  \citet{1994ApJS...95....1E} has seen extensive utilisation in many investigations.
Templates based on more recent observations, such as those of \citet{2006ApJS..166..470R}, have further parametrized the intra-population dispersion of quasar SEDs but still focus on extended frequency coverage at relatively low resolution. Such templates will continue to see extensive use but they have only limited utility in the context of investigations using precision optical and near-infrared photometry (as will be provided by LSST and \textit{Euclid}) to study the AGN and quasar populations over large ranges of redshift and luminosity. 
The majority of quasars possess prominent emission features in the rest-frame ultraviolet and optical and a significant subset ($\simeq 20$ per cent; \citealt{2003AJ....125.1784H}) also show strong absorption features. While optical and near-infrared photometric passbands used for large surveys are normally relatively broad, the majority deliberately possess sharp short- and long-wavelength transmission cutoffs. As a consequence, making accurate (few per cent) predictions for the colours of quasars as a function of redshift requires model SEDs with a resolution of at least $R\approx 500$. 

A different approach, which allows investigation across
the extended rest-frame wavelength range probed by a single observed-frame photometric passband  over a significant redshift range, is the use of templates based on composite quasar spectra.
One such template which is highly cited is the composite presented by
\citet{VandenBerk01}.
However, such composite spectra  are generally derived from samples with a large range in luminosity as a function of rest-frame wavelength, which means that different parts of the template are appropriate for different regions of luminosity-redshift space.
In the case of the Vanden Berk composite, there is a significant contribution to the SED at wavelengths longer than $\approx$4000\,\AA \ from host galaxy light associated with the low-redshift AGN used to generate the rest-frame optical composite spectrum. The galaxy contribution can make up $\approx$40\, per cent of the composite SED at long wavelengths.
At the same time, the rest-frame ultraviolet region of the composite is derived from higher-redshift quasars which are significantly more luminous. It is therefore desirable to develop model SEDs which allow the luminosity and redshift to change as independent input parameters.

\subsection{Previous parametric models}
\subsubsection{Early  quasar SED models}

The use of parametric models to help develop selection algorithms and calculate the completeness of quasar samples began with searches for (then) high-redshift quasars of $z\approx4$ \citep{1991ApJS...76....1W}. At such redshifts, quasars were intrinsically rare and candidate selection was made challenging due to the presence of much more common contaminant populations that possessed similar photometric properties. Comparable models \citep{1999AJ....117.2528F, 2001AJ....121...31F} were used to calculate the completeness of the initial SDSS quasar sample \citep{2002AJ....123.2945R} and extend the high-redshift searches beyond redshift six. All these models focused on predicting the observed-frame colours for wavelengths within the extended `optical' wavelength interval, $\approx$3500-10\,000\,\AA, where relatively deep broad-band photometric observations were possible. A similar model was used in an early investigation of photometric redshifts for quasars by \citet{2000A&A...359....9H}.

The availability of wide-field near-infrared surveys, including the Two-Micron All Sky Survey  \citep[2MASS;][]{2006AJ....131.1163S}, UKIRT Infrared Deep Sky Survey  \citep[UKIDSS;][]{Lawrence07} and, more recently, \textit{WISE} \citep{Wright10} extended the accessible wavelength range for relatively faint quasars, first to 2.5\,$\mu$m and with \textit{WISE} to 4.5\,$\mu$m. Parametric models were extended to rest-frame wavelengths of several microns, including contributions to the quasar SED from hot dust \citep{Maddox08}. Such models provide predictions of the optical through near-infrared colours of an accuracy sufficient to allow the estimation of photometric redshifts. Of particular note was the parametrization of the quasar `continuum' longward of $\approx$10\,000\,\AA. Observations of individual quasars \citep{Glikman2006} and near-infrared photometry of luminous quasars \citep{Maddox06, Maddox08} resulted in the identification of a component that can be reproduced by emission from a blackbody with temperature $\approx$1200\,K \citep[see][for a recent investigation]{2021MNRAS.501.3061T}. 

\subsubsection{Current state of the art}

The two parametric models that have seen considerable use in the prosecution of high-redshift quasar searches are those developed by Hewett \citep{Maddox08, Maddox12} and McGreer and Fan \citep{1999AJ....117.2528F, 2013ApJ...768..105M}. The former model was a key element in the probabilistic scheme developed by \citet{2012MNRAS.419..390M} that led to the discovery of the first quasar to be observed with a redshift exceeding seven \citep{2011Natur.474..616M}, and also formed the basis for the SED-fitting scheme used in the selection of $z>6$ quasars from the Dark Energy Survey and VISTA Hemisphere Survey by \citet{2017MNRAS.468.4702R, 2019MNRAS.487.1874R}. Comparison of the predictions of the two models for `mean' high luminosity quasars at significant redshifts ($z\gtrsim 2$) up to redshift $z\approx 8$ show very good agreement (Hewett and McGreer, private communication). The model parametrizations are though quite different. Taking the emission lines for example; the Hewett model uses a template quasar spectrum derived from the \cite{Francis91} composite spectrum, extended longward of the H$\alpha$ line, with just two scaling parameters to set the strength of the emission line spectrum and (separately) the H$\alpha$ line, whereas the McGreer model includes the facility to specify the strengths and velocity-widths of 60 individual emission lines.

\subsection{Improvements presented in this work}

\begin{figure}
    \centering
    \includegraphics[width=\columnwidth]{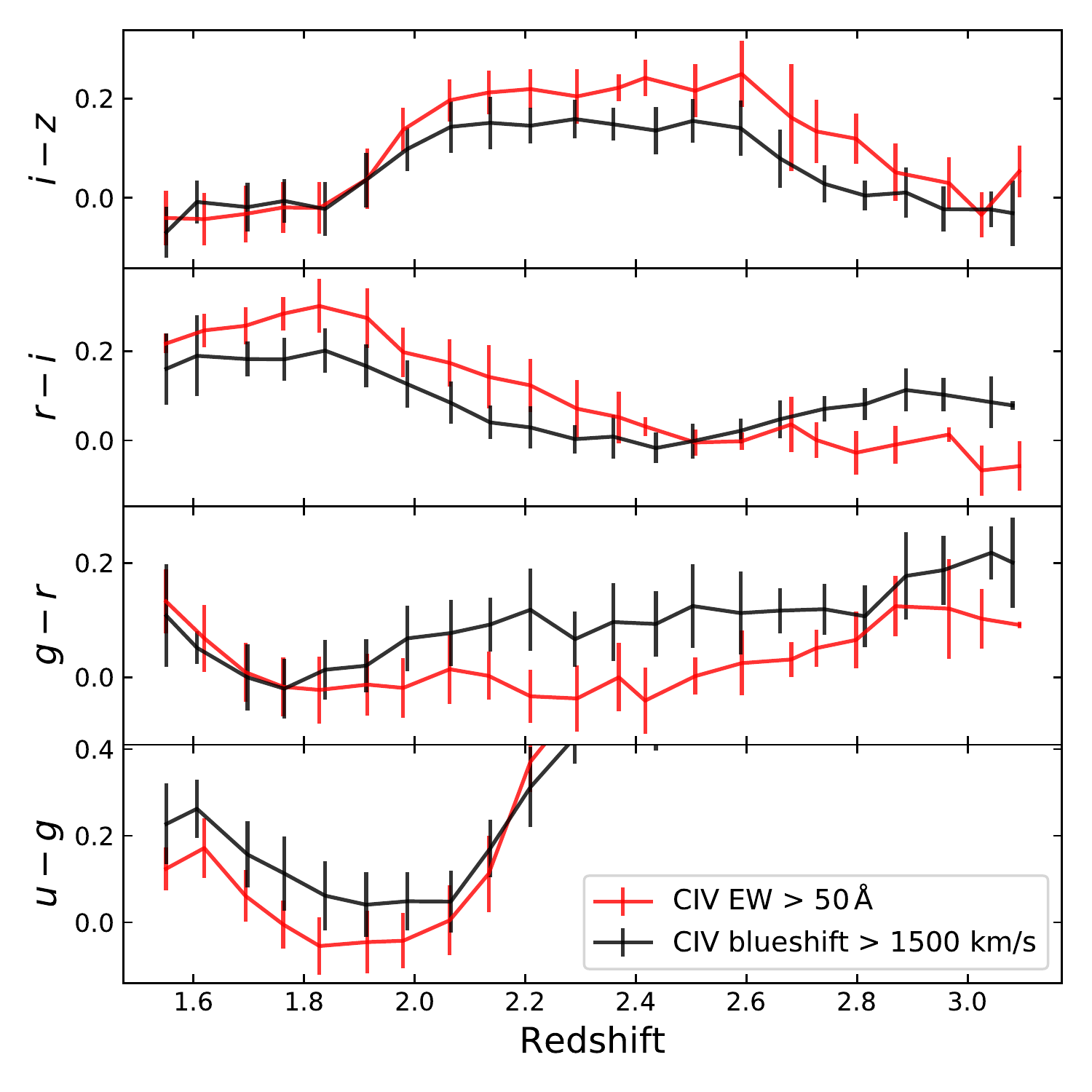}
    \caption{Median SDSS \textit{ugriz} colours as a function of redshift.
    In red: objects with high equivalent width \civ\ $\lambda$1549 emission also have stronger Ly$\alpha$ emission, leading to bluer \textit{u-g} and \textit{g-r} colours when compared to objects with weaker, highly blueshifted line emission (in black).
    At redshifts $z>2.3$, the Lyman break moves through \textit{u}.
    Both samples have $18.6<i_{\rm AB}<19.1$, and exclude BAL quasars.
    \civ\ properties are taken from \citet{Rankine20}.
    Error bars show the standard deviation within each bin.
    Even when matching in luminosity, the photometric colours can change by $\sim$0.1\,mag for quasars with different emission-line properties.}
    \label{fig:CIV_extrema}
\end{figure}

\subsubsection{Emission line properties}
\label{sec:beff}

Understanding, at least from an observational perspective, of the range of ultraviolet and optical SEDs of quasars has improved significantly over the last 20 years. Of particular relevance are the systematic changes in the equivalent width and kinematics of the strongest emission lines \citep[e.g.][]{2002MNRAS.337..275C, 2002ApJ...566L..71S, 2007ApJ...666..757S, Richards11, 2016ApJ...833..199J, Rankine20, 2020MNRAS.496.2565T, 2021MNRAS.505.3247T}. A significant difference in the model presented here is the ability to incorporate the systematic emission-line diversity using just a small number of parameters. 

The emission line properties of quasars are fundamentally linked to the ionising SED \citep{ 2011AJ....142..130K, Krawczyk13, 2015AJ....149..203K}. 
It has been proposed by \citet{1977ApJ...214..679B} that the equivalent width of strong emission lines in quasar spectra is (anti-)correlated with the intrinsic luminosity of the source.
However, this correlation has also been ascribed as spurious and due to selection effects, with contradictory results in the literature for different samples \citep[see section 3.1 of][for a review]{Sulentic00}.
In reality, the emission line properties are expected to be driven by the accretion rate (i.e. Eddington fraction) of the AGN, with a secondary dependence on the black hole mass \citep{2004ApJ...617..171B, 2019A&A...630A..94G}.
As one looks to higher redshifts, the intrinsic luminosity of a source at any given apparent magnitude increases, and so for any flux-limited sample of AGN, the average Eddington rate will increase as we move to higher redshifts.
Such sources are known to generally display weaker emission lines, with the high-ionization lines in particular showing evidence for outflows through asymmetric and blueshifted emission (see fig.~5 of \citealt{2021MNRAS.505.3247T}).
In Fig.~\ref{fig:CIV_extrema}, we show how these differences can lead to changes in colour of 0.1\,mag or more, even when matching in luminosity. 

\subsubsection{Host galaxy contribution}

A second major improvement relates to the determination of host galaxy contributions to the `quasar' SED and the ability to isolate the pure quasar SED. The Hewett-model was developed using the luminous, $i_{\rm AB} \le 19.1$,  SDSS DR7 quasars to determine the parameters but it is now possible to utilise the much larger sample of quasars included in SDSS DR16. Importantly, the DR16 quasar sample probes a significantly greater dynamic range in luminosity at fixed redshift. Mid-infrared photometry from \textit{WISE}, which provides much improved constraints on the rest-frame $\approx$1$\mu$m SED, where the relative contribution to the quasar+host SED is maximal, is now also available. As a result, it is possible to determine the relationship between quasar and host galaxy as a function of luminosity far more effectively. 

\subsection{Structure of this paper}

\begin{figure*}
    \centering
    \includegraphics[width=2\columnwidth]{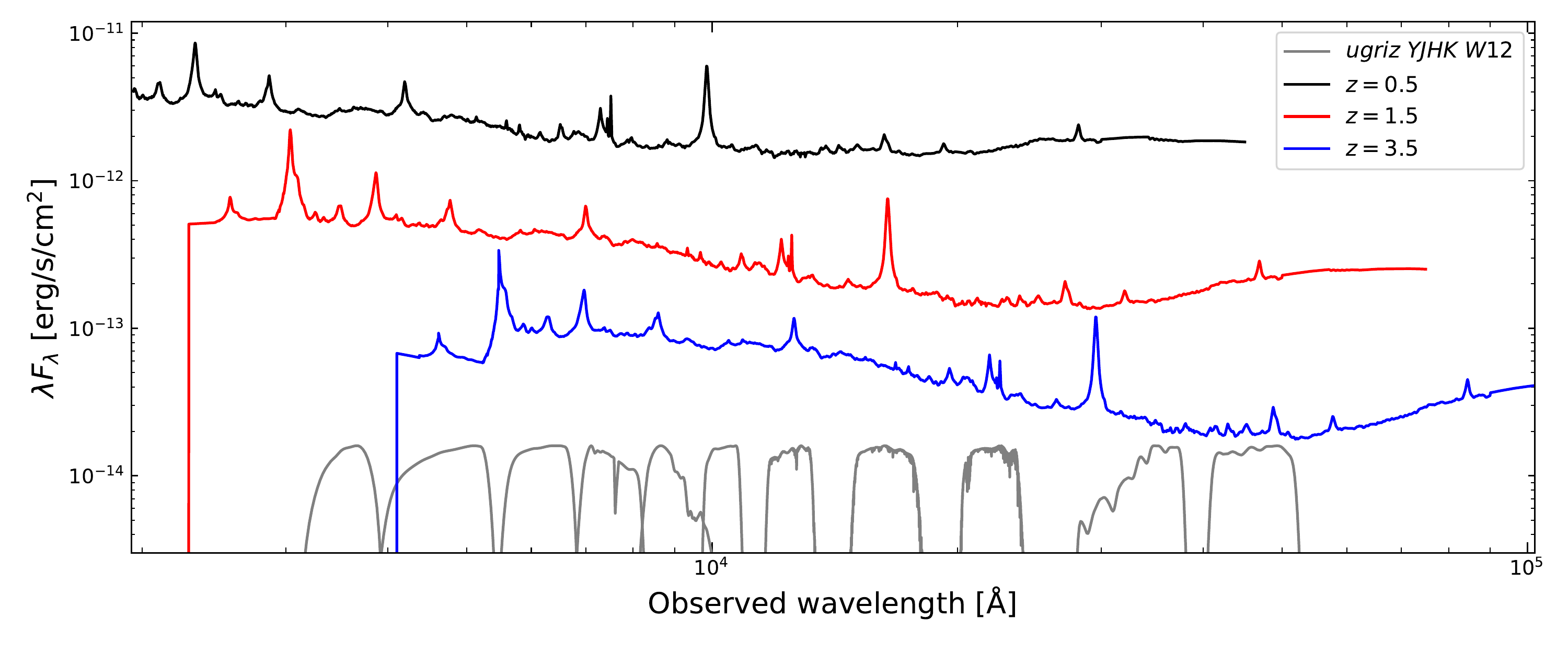}
    \caption{Examples of our model SED in the observed-frame, for three different redshifts. SDSS \textit{ugriz}, UKIDSS \textit{YJHK} and \textit{WISE} \textit{W12} filter response curves are shown for comparison.
    Higher-redshift quasars tend to be more luminous and have less contamination from starlight in their host galaxies, resulting in a sharper minimum at $\simeq$1\,\micron\ in the rest-frame. The effect of Lyman suppression arising from incomplete transmission through the inter-galactic medium can be seen bluewards of Ly$\alpha$ in the $z=3.5$ model.}
    \label{fig:model}
\end{figure*}

In this work, we present a parametric model for the rest-frame 912\,\AA\ to 3\,\micron\ region of the quasar SED, which corresponds to the observed-frame optical and near-infrared colours.
The quasar SED model builds on the model used by \citet{2006MNRAS.367..454H} and \citet{Maddox08, Maddox12},
incorporating the improvements outlined above, 
and is made publicly available for the first time as a {\sc python} code.\footnote{\url{https://github.com/MJTemple/qsogen/}}
Three examples of our model SED are shown in Fig.~\ref{fig:model}.

The model which we herein present is capable of reproducing the observed-frame optical and near-infrared colours of the majority of quasars, to a high degree of accuracy, over redshifts $0.2 \lesssim z \lesssim 7.5$. Given the SED properties, the model could be used to simulate the appearance of the quasar population in photometric surveys using either probability density functions for key parameters \citep[e.g.][]{2012MNRAS.419..390M} or a monte-carlo sampling scheme \citep[e.g.][]{2013ApJ...768..105M}. The relatively small number of parameters needed to reproduce the properties of individual quasars means that the parameter distribution(s) for observed quasar populations can be determined though fits of the model to photometry of individual quasars. Where redshifts are not available, model fits for individual objects, with redshift as an additional free parameter, could also be used to generate photometric redshifts. 

The structure of the model is described in Section~\ref{sec:model}.
By cross-matching the latest SDSS quasar catalogue to UKIDSS and \textit{WISE}, we obtain a large sample of quasars with photometric data covering the \textit{ugrizYJHKW12} bands, which we use to calibrate our model parameters in Section~\ref{sec:params}.
In Section~\ref{sec:verif}, we show that our model provides good predictions for populations which are significantly fainter than the sample which has been used to calibrate the model, and also for populations at much higher redshift. Our predictions for the average quasar colours which will be seen in LSST and \textit{Euclid} are given as a function of redshift and flux in Section~\ref{sec:predict}, and we compare our model to previous work in Section~\ref{sec:discuss}.

All SDSS magnitudes and colours in this work are reported on the AB system \citep{1983ApJ...266..713O}.
We report near-infrared UKIDSS and VIKING and mid-infrared \textit{WISE} magnitudes and colours on the Vega system, the native magnitude system for those surveys, where Vega is assumed to have zero magnitude in all passbands.
Predicted LSST and \textit{Euclid} colours are reported on the AB system.
We assume a flat $\Lambda$CDM cosmology with $\Omega_m=0.27$, $\Omega_{\Lambda}=0.73$, and $\textrm{H}_0=71 \kmpspMpc$. All emission lines are identified with their wavelengths in vacuum in units of \AA ngstr\"{o}ms.

\section{The Parametric SED Model}
\label{sec:model}

We begin by constructing a parametric SED model to describe quasar emission in the 912\,\AA\ to 3\,\micron\ rest-frame wavelength range.
In Section~\ref{sec:params} the model parameters are then 
calibrated using the median observed \textit{ugrizYJHKW12} colours of the SDSS DR16 quasar population.

\subsection{Ultraviolet--optical continuum}

The 900\,\AA\ to 1\,\micron\ region of a type-1 quasar SED is dominated by a blue continuum,
corresponding to the low-frequency tail of the direct emission from the accretion disc.
We characterise this continuum using a continuous broken power-law, with both the slopes, $\alpha_{1,2}$, and position of
the break,  $\lambda_\textrm{break}$, able to vary. 
\begin{equation}
\begin{split}
    f_\nu  \propto & \begin{cases}
     \nu^{\alpha_1} \hspace{5mm} & \nu > c/\lambda_\textrm{break} \\
     \nu^{\alpha_2} & \nu < c/\lambda_\textrm{break}
    \end{cases} \\
\end{split}
\end{equation}
equivalently
\begin{equation}
\begin{split}
    f_\lambda  \propto & \begin{cases}
     \lambda^{-\alpha_1-2} \hspace{5mm}& \lambda < \lambda_\textrm{break} \\
     \lambda^{-\alpha_2-2} & \lambda > \lambda_\textrm{break}
    \end{cases} \\
\end{split}
\end{equation}

At wavelengths
$\lambda\leq1200$\,\AA, the power-law slope is modified to become $\alpha_3 = \alpha_1 -1$ to account for the sharp change in the continuum observed in quasar spectra \citep[][]{1980ApJ...239..483G}.

\subsection{Hot dust}

Near-infrared emission at wavelengths $1<\lambda<3\,\micron$ is dominated by emission from hot dust, which we characterise using a single blackbody, as described by \citet[][ see  Appendix~\ref{sec:HD}]{2021MNRAS.501.3061T}.
The temperature, $T_\textrm{BB}$, and normalisation relative to the power-law continuum at 2\,\micron, \texttt{bbnorm},
of this blackbody are free to vary.
\begin{equation}
    f_\lambda  = \texttt{bbnorm}  \times B_\lambda(T_\textrm{BB}) \times
    \left( \frac{\displaystyle f_{\lambda, \textrm{ power-law}}
    }{B_\lambda (T_\textrm{BB})}\right)_{\lambda= 2\small{\micron}}
\end{equation}
where $B_\lambda(T)$ is the Planck function:
\begin{equation}
    B_\lambda(T) = 
    \frac{2 h c^2}{\lambda^5}
    \frac{1}{e^{hc/\lambda k_B T} - 1}
\end{equation}
\subsection{Balmer continuum}
We use the prescription from \citet{Grandi82} to describe the broad $\sim$3000\,\AA\ `bump' due to blended emission from high-order Balmer lines, optically thin Balmer continuum, and two-photon emission from neutral Hydrogen:
\begin{equation}
\begin{split}
    f_\lambda & = f_0 \times  B_\lambda(T_\textrm{BC})(1-e^{-\tau_\lambda});
        \hspace{5mm} \lambda<\lambda_\textrm{BE} \\
    \tau_\lambda & = \left(\frac{\lambda}{\lambda_\textrm{BE}}\right)^3 \\
    f_0 & =\texttt{bcnorm} \times 
    \left(\frac{
    \displaystyle f_{\lambda, \textrm{ power-law}}
   }{B_\lambda(T_\textrm{BC})(1-e^{-\tau_\lambda})}\right)_{\lambda= 3000\textrm{\AA}}
\end{split}
\end{equation}
where $T_\textrm{BC}=15\,000$\,K is the electron temperature of the Balmer continuum, $\lambda_\textrm{BE}=3646$\,\AA\ is the wavelength of the Balmer edge, and $\tau_\lambda$ is the optical depth of the Balmer continuum, fixed such that the continuum is optically thick at the edge, i.e.\,$\tau_{\lambda_\textrm{BE}}=1$. This function is then broadened via convolution with a Gaussian, with a full width at half maximum  of 5000\kms, to approximate the kinematics of the Balmer-emitting gas in broad line AGN. The normalisation relative to the power-law continuum at 3000\,\AA, \texttt{bcnorm}, is the only parameter free to vary.

%At this stage the model can be normalised at either 3000 or 5100\,\AA, for a given monochromatic luminosity.

\subsection{Emission lines}
\label{sec:lines}

\begin{figure}
    \centering
    \includegraphics[width=\columnwidth]{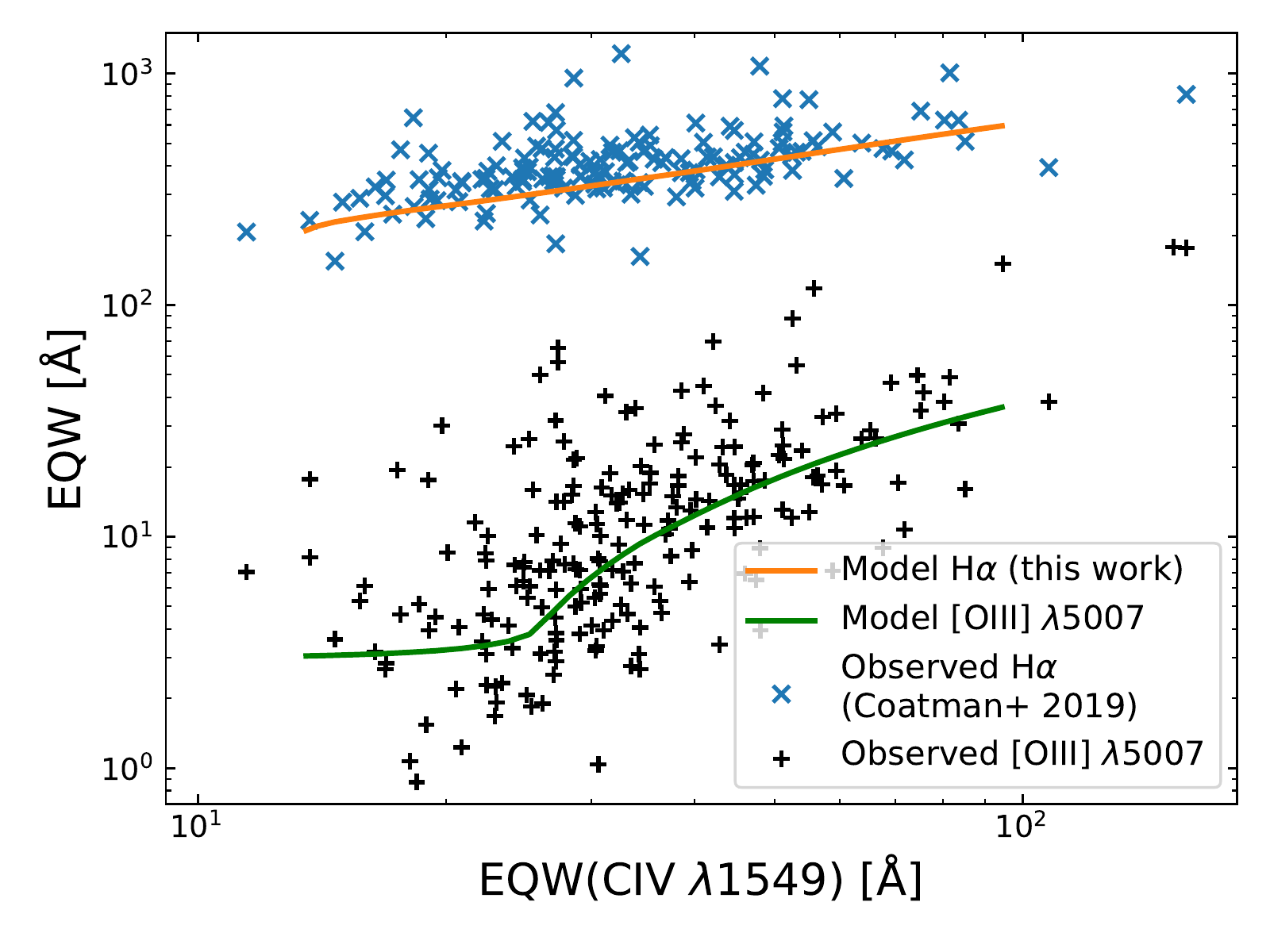}
    \caption{The equivalent widths of the rest-frame ultraviolet \ion{C}{IV}\,$\lambda1549$ emission line \citep{Rankine20}, and the rest-frame optical  [\ion{O}{III}]\,$\lambda5007$ and H$\alpha\,\lambda6565$ lines \citep{Coatman17, Coatman19}. Crosses show measurements in individual objects; the strengths of both broad and narrow optical lines display positive correlations with the strength of \ion{C}{IV} emission.
    The solid lines show how our SED model reproduces these trends by using a single parameter to describe the emission-line properties across the ultraviolet-optical wavelength range.}
    \label{fig:line_eqws}
\end{figure}

\begin{figure*}
    \centering
    \includegraphics[width=2\columnwidth]{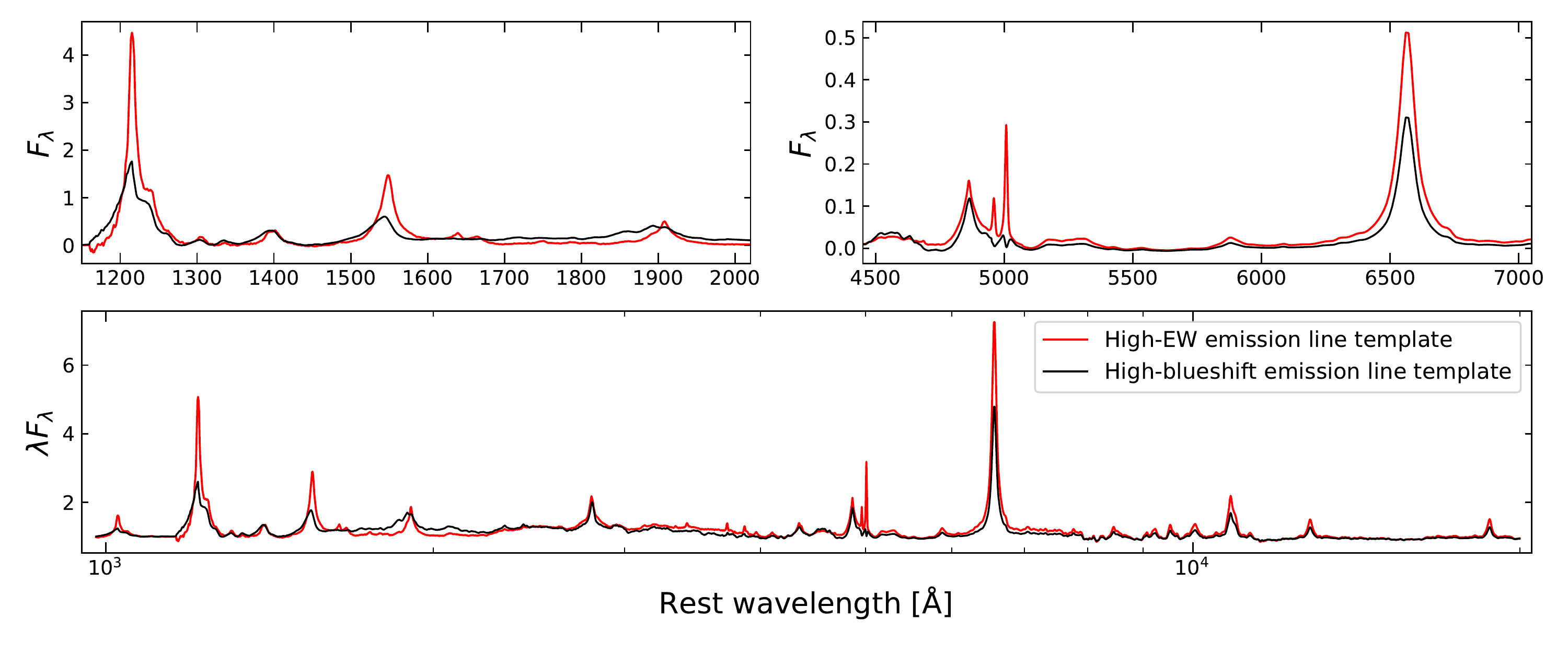}
    \caption{The high blueshift and high equivalent width emission line templates implemented in the quasar SED model.
    Quasars with higher luminosities and larger Eddington fractions are known to be more likely to display emission lines which are weaker and display larger shifts to the blue in high-ionization ultraviolet transitions \citep{2021MNRAS.505.3247T}.
    Such objects are known to show weaker optical line emission in both narrow [\ion{O}{III}] and the broad Balmer series (Fig.~\ref{fig:line_eqws}).
    }
    \label{fig:emlines}
\end{figure*}

It is now well established that the properties of the various ultraviolet emission lines observed in luminous quasars are highly correlated. 
Figs.~11 and 12 of \citet{Richards11}, fig.~A2 of \citet{Rankine20} and fig.~5 of \citet{2021MNRAS.505.3247T} show examples of how systematic changes in the 1000-2000\,{\AA} region of the spectrum can be inferred using the properties of just one emission line.
Further work by \citet{Coatman17, Coatman19} has shown that the  rest-frame optical emission line properties  also correlate with the ultraviolet emission (Fig.~{\ref{fig:line_eqws}}).
More recently, \citet{2021MNRAS.504.5556T} and Rivera et al. (in prep) have demonstrated that the trends in the ultraviolet are strongly correlated with the X-ray emission, specifically the parameter $\alpha_\textrm{ox}$, suggesting that the correlated diversity in the emission-line properties is driven by changes in the shape of the SED responsible for ionizing the line-emitting regions.

For this work, we seek to reproduce the median quasar colours across a wide range of redshifts and intrinsic luminosities using as few parameters as possible.
We therefore construct two emission line templates, which represent the extrema of the range of observed quasar SEDs (Fig.~\ref{fig:emlines}).
The construction of these templates is described in detail in Appendix~{\ref{sec:emlines}}.
In terms of the equivalent width of the high-ionization ultraviolet lines such as {\civ},
these two templates represent very strong and very weak line emission.
By interpolating between these two extremes,
it is possible to reproduce changes in line strength to the level of precision required when modelling broad-band photometric data.
Moreover, by linking this interpolation to the intrinsic
luminosity of the quasar model,
such that brighter objects have smaller line equivalent widths,
we naturally produce a Baldwin effect (Section~\ref{sec:beff}).

To implement this effect, we introduce a parameter \texttt{emline\_type} which allows for interpolation between the two extreme templates. The value of this parameter at any given redshift is controlled by the average absolute magnitude $M_i$ at that redshift together with the multiplicative factor \texttt{beslope}:
\begin{equation}
    \texttt{emline\_type} = \texttt{beslope} \times (M_i(z) + 27)
\end{equation}
where the zeropoint is chosen to match the average absolute magnitude $M_i = -27$ for the median $\lambda$1908 complex at $z\simeq2$.

When calibrating our model parameters in Section~{\ref{sec:params}},
we find that \texttt{beslope}, i.e. the slope of the anti-correlation between quasar luminosity and emission line strength,
can be constrained using the SDSS, UKIDSS and unWISE photometry.
In other words, we infer the existence of a Baldwin effect directly from the photometric data, without using any spectroscopic knowledge of the line strengths in our quasar sample.

Within the SED model, the equivalent width of each line in the template is preserved as the shape of the continuum changes, subject to an overall scaling factor \texttt{scal\_em}.
We found that preserving the emission line equivalent widths resulted in a better representation of the colours, but the model code also has the ability to instead preserve the relative emission line fluxes.
Our model code also has the capability to independently re-scale the narrow optical emission lines, or the broad H$\alpha$ or Ly$\alpha$ emission lines. Such functionality is not necessary when investigating the average colours of the quasar population, and so we do not use these code parameters in this paper.

\subsection{Host galaxy flux}

We include the facility to incorporate emission from host galaxies.
For the luminous quasars we consider, the primary effect of starlight from the host galaxy is to add flux at and around the 1\,\micron\ minimum in the total quasar SED. For optically unobscured, blue quasars such as those from SDSS, the strength of emission from any young stellar population is very hard to quantify as the quasar power-law continuum will be much brighter than the host galaxy at shorter wavelengths. It is therefore hard to quantify any change in shape of the host galaxy SED across the luminosity and redshift range we consider.
We use an S0 template from the SWIRE library \citep{ 2007ApJ...663...81P, SWIRE}, having found that this choice of template provides the best fit to the SDSS-UKIDSS-unWISE colours of known quasars in Section~\ref{sec:params}. The galaxy contribution to the median quasar plus host SED will be a combination of a significant fraction of quasars with hosts dominated by old stellar populations and other objects with hosts that have experienced different star-formation histories. The selection of the S0 host is thus not surprising.

Modelling the spectra and colours of individual quasars and the population of fainter AGNs will necessarily involve incorporating different host galaxy SEDs. Our model code is capable of utilising any specified galaxy SED and a specific application is the estimation of photometric redshifts for quasars and AGN utilising a range of host-galaxy SEDs.
\citet{2019NatAs...3..212S} provide a recent review of template-based and other approaches to the estimation of photometric redshifts for galaxies including those with contributions from AGN light.
Here, however, we focus on reproducing the median colours of the SDSS quasar population and employ a single host-galaxy SED. 

The strength of host galaxy emission is incorporated into the model with two parameters: \texttt{fragal}, the fraction of total flux in the wavelength region 4000-5000\,\AA\ due to the galaxy for an object at a reference luminosity $L_0$, and \texttt{gplind}, the power-law index controlling how the luminosity of the galaxy component changes as a function of the quasar luminosity:
\begin{equation}
    \frac{L_\textrm{galaxy}}{L_0} = \frac{\texttt{fragal}}{1-\texttt{fragal}} 
    \left(\frac{L_\textrm{quasar}}{L_0}\right)^\texttt{gplind}
    \label{eq:galnorm}
\end{equation}
For values of \texttt{gplind} which lie between 0 and 1, the host galaxy component gets brighter as the quasar gets brighter (e.g. as one moves to higher redshifts within a flux-limited sample), but the fractional contribution of the galaxy to the total flux decreases.

The luminosities used are the absolute \textit{i}-band magnitudes at redshift 2, derived using the \textit{K}-correction from \citet{2006ApJS..166..470R} and as reported in the \citet{DR16Q} catalogue. 
For each flux cut that is used, the median $M_i$ is calculated in each redshift bin, giving an empirical redshift-luminosity relation for each sample which is then assumed when calculating the galaxy normalisation at any given redshift (see Fig.~\ref{fig:zlum_lumval}).
The reference luminosity $L_0$ is taken to be $M_i=-23$. This is close to the average luminosity of SDSS DR16 quasars with $18.6<i_\textrm{AB}<19.1$ at $z=0.35$.

\subsection{Dust reddening}

\begin{figure}
    \centering
    \includegraphics[width=\columnwidth]{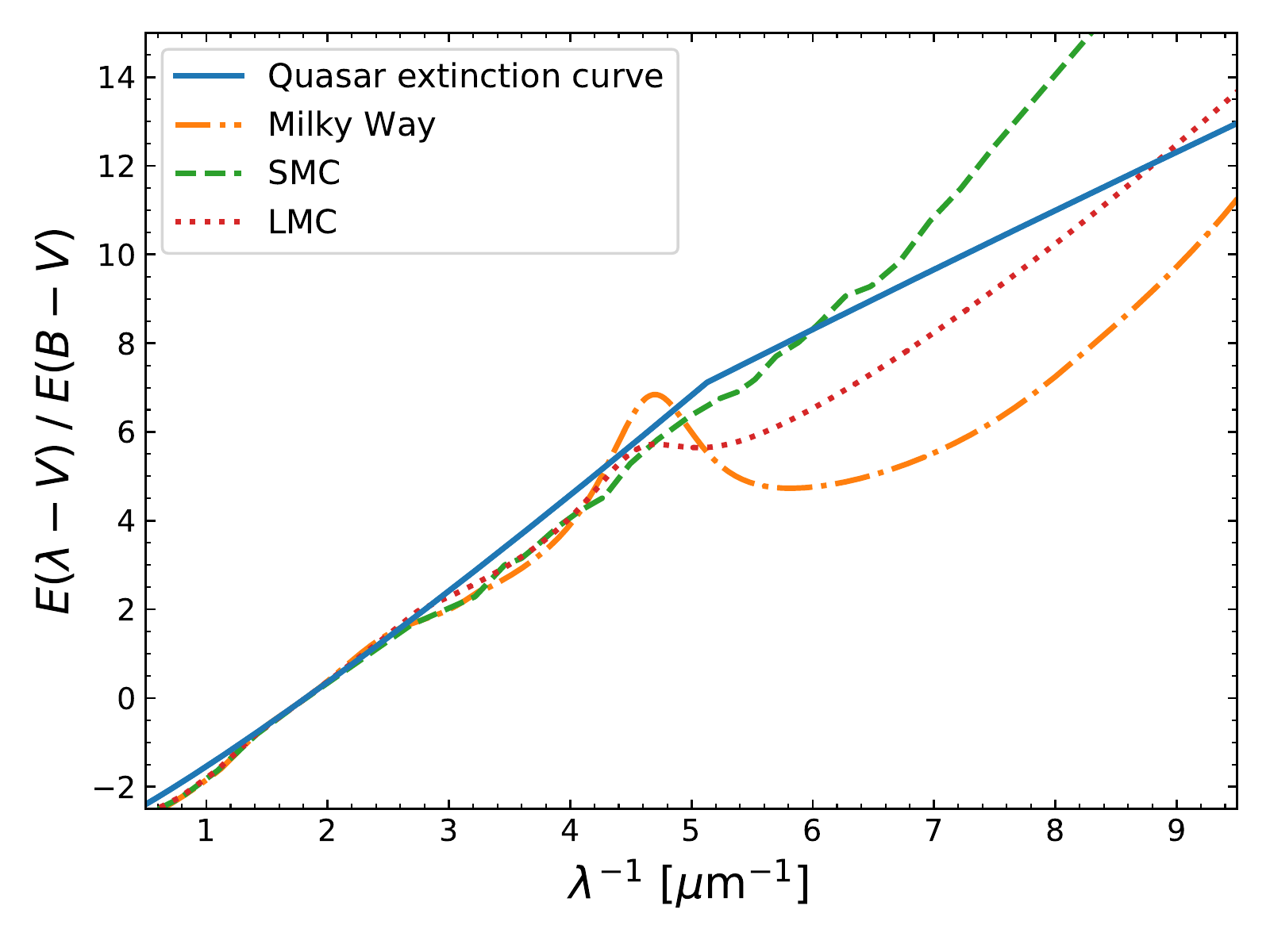}
    \caption{The quasar extinction curve included in the model, compared to commonly-used curves from our own Galaxy and the Small and Large Magellanic Clouds. The quasar extinction curve has no feature at 2200\,\AA\ ($\lambda^{-1}=4.5\,\mu\textrm{m}^{-1}$), and rises less steeply than the SMC extinction curve at wavelengths shorter than 1700\,\AA\ ($\lambda^{-1} > 6\,\mu\textrm{m}^{-1}$).}
    \label{fig:ext_curve}
\end{figure}

Attenuation due to dust at the redshift of the quasar using an empirically derived extinction curve can be incorporated in the model. The user can specify the form of the extinction curve by providing an ascii file specifying the value of $E(\lambda-V)/E(B-V)$ as a function of wavelength $\lambda$. Our own specification of the extinction curve appropriate for quasar samples (Fig.~\ref{fig:ext_curve}) is somewhat similar to those presented by \citet{2004MNRAS.348L..54C} and \citet{2010A&A...523A..85G}. The curve rises more steeply with decreasing wavelength than the Milky Way curve, has no 2200\,\AA \ feature and is similar to the extinction curve of the Small Magellanic Cloud (SMC) for wavelengths $\gtrsim$1700\,\AA. Shortward of 1700\,\AA, however, the extinction increases significantly less rapidly than for the SMC or, indeed, the LMC and Milky Way curves. In other words the form of the curve shortward of $\approx$1700\,\AA \ is greyer than observed in the Local Group. 

The extinction curve was calculated using \textit{i}- and \textit{K}-band photometry of SDSS DR7 quasars with $2.0 < z < 3.0$ to calculate the rest-frame optical (3000-6500\,\AA) reddening, $E(B-V)$.
Composite SDSS quasar spectra for samples with different optical reddenings covering the rest-frame wavelengths 1100-2500\,\AA\ then enabled the ultraviolet extinction curve to be defined. While not explicitly published previously the extinction curve has been used extensively in our earlier work \citep[e.g.][]{Maddox08, Maddox12, 2011MNRAS.410..860A, Banerji12, Banerji13, Banerji15, Wethers18, Temple19}.
Adopting an extinction curve as steep as local examples at $\lambda < 1700$\,\AA \ would be inconsistent with numerous of our previous results. Figure 22 of \citet{2011MNRAS.410..860A} provides an example, where the extinction curve provides an excellent match to the reddening of broad absorption line quasars over the redshift range $1.7 < z < 4.0$.

The quasar flux, excluding the host galaxy component, can be reddened using this quasar extinction curve with the $E(B-V)$ as a free parameter.
The colour distribution of SDSS quasars is largely Gaussian, with only a small tail due to reddening \citep{ 2003AJ....126.1131R, Hopkins04}.
We apply an iterative sigma-clipping procedure to exclude this tail when calculating the median colour in any given redshift bin. The observed colours we report are therefore assumed to be not significantly affected by extinction, and the $E(B-V)$ is  held fixed at zero when calibrating  the other model parameters.

\subsection{Lyman-absorption suppression}

We have the ability to include a Lyman-limit system (LLS), which has the effect of setting the model flux to zero at all wavelengths below some cut-off. 
The default cut-off wavelength is $\lambda_\textrm{LLS}=912$\,\AA.

At redshifts $z>1.4$, flux at $\lambda < 1216$\,\AA\ is suppressed to account for the incomplete transmission through the inter-galactic medium  of flux shortwards of \lya.
The effect of this suppression on the average colours of a sample of quasars is relatively well-understood, however we note that individual objects can display very different amounts of Lyman transmission depending on their individual sight-lines.
We use the prescription of \citet{2013MNRAS.430.2067B}, which has been calibrated using the results of \citet{2008ApJ...681..831F}:
\begin{equation}
    \tau_\textrm{eff, \lya}(z) = 0.751\left(\frac{1+z}{1+3.5}\right)^{2.90}-0.132 
\end{equation}
The suppression due to \lyb\ and \lyg\ is included, assuming the theoretical ratio of oscillator strengths \citep[e.g.][]{2020MNRAS.497..906K}:
\begin{equation}
\begin{split}
\tau_\textrm{eff, \lyb}(z) &= 0.160\times\tau_\textrm{eff, \lya}(z) \\
\tau_\textrm{eff, \lyg}(z) &= 0.056\times\tau_\textrm{eff, \lya}(z)
\end{split}
\end{equation}

Finally, the SED is redshifted into the observed-frame and multiplied by the passband filter responses to obtain synthetic photometry.

\section{Calibrating the model parameters}
\label{sec:params}

In the previous section, we described the construction of a parametric quasar SED model. 
Before using this model to provide predictions for future surveys, in this section we first calibrate the model parameters with 
the average colours of known quasar populations which have been binned in redshift and flux.

\subsection{Data}
\label{sec:photometry}

We make use of the SDSS sixteenth data release (DR16) quasar catalogue, the selection of which is summarised in \citet{DR16Q}. As well as being the final data release from SDSS-IV, the DR16 catalogue constitutes the largest sample of spectroscopically confirmed quasars currently available.
We use the `primary' redshifts from the DR16 catalogue, although we take care to exclude outliers when computing the median colour in each redshift bin (Section~\ref{sec:binning}), and our results would be unchanged if we instead used the `PCA' redshift.

Starting with the 750\,414 objects in the DR16 `quasar-only catalogue' \citep{DR16Q}, 
we exclude 82 objects which have been identified by \citet{Flesch2020} as being non-quasar contaminants.
The number of quasars in the catalogue with redshifts $z>5$ is very low, and their pipeline redshifts are highly uncertain, so we also exclude these objects from the catalogue.
Finally, we exclude objects with Galactic extinction $E(B-V)>0.3$.
This yields a total of 748\,620 quasars with redshifts $0<z<5$.
%A subset of the DR16 quasars, originally selected and presented in the seventh data release  \citep[DR7,][]{2010AJ....139.2360S}, have a uniform flux limit of $i<19.1$ below redshift 3 and a slightly fainter limit above. 

\subsubsection{SDSS photometry}

We use \textit{ugriz} point spread function magnitudes from the SDSS quasar catalogue.
SDSS reports magnitudes on the `asinh' system \citep{Lupton99}. 
We only use objects with $i_\textrm{AB}<20.6$, where the majority of SDSS quasars have  $z_\textrm{AB}<20.3$ and their asinh magnitudes and logarithmic Pogson magnitudes differ by less than one per cent in flux.
We also exclude \textit{u} at redshifts $z>2.3$ and \textit{g} at $z>3.0$ where quasars have little flux as they `drop-out' of these bands.
The asinh magnitude is therefore well-approximated by the Pogson magnitude and SDSS zeropoints are converted to the AB standard using
\begin{equation}
    \begin{split}
        u_\textrm{AB} = u_\textrm{SDSS} - 0.04 \\
        z_\textrm{AB} = z_\textrm{SDSS} + 0.02
    \end{split}
\end{equation}
The SDSS \textit{gri} zeropoints are believed to match the AB system.\footnote{\url{https://www.sdss.org/dr16/algorithms/fluxcal\#SDSStoAB}}

%AB magnitudes are converted to the Vega system using 
%\begin{equation}
%    \begin{split}
%        u_\textrm{Vega} = u_\textrm{AB} - 0.913 \\
%        g_\textrm{Vega} = g_\textrm{AB} + 0.081 \\
%        r_\textrm{Vega} = r_\textrm{AB} - 0.169 \\
%        i_\textrm{Vega} = i_\textrm{AB} - 0.383 \\
%        z_\textrm{Vega} = z_\textrm{AB} - 0.542
%    \end{split}
%\end{equation}

\subsubsection{WISE photometry}

\begin{figure}
    \centering
    \includegraphics[width=\columnwidth]{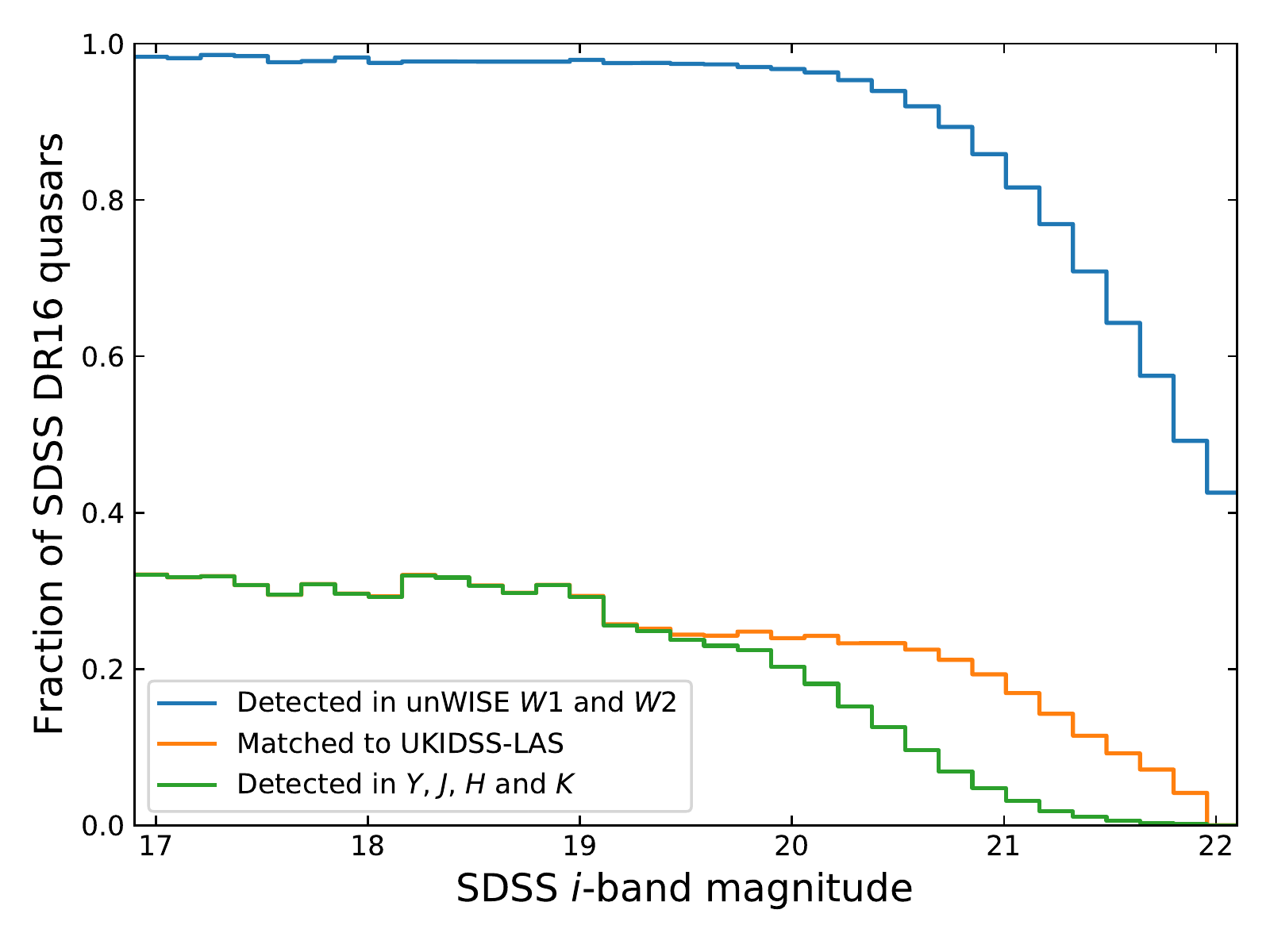}
    \caption{The completeness of the SDSS DR16 quasar catalogue in UKIDSS-LAS DR11plus and unWISE. 
    unWISE covers the whole sky, and provides \textit{W1} and \textit{W2} data for 98 per cent of DR16 quasars with $i<20.1$.
    UKIDSS-LAS covers around three tenths of the SDSS footprint. Within that area, almost all DR16 quasars with $i_\textrm{AB}<19.1$ are detected in all four \textit{YJHK} bands, but the number of matches begins to drop above $i\simeq19.5$.}
    \label{fig:completeness}
\end{figure}

We cross-match the SDSS DR16 quasar catalogue to the deepest source of mid-infrared \textit{WISE}  \citep{Wright10} data which is currently available:  the unWISE catalogue presented by \citet{Schlafly19}, which makes use of the  \citet{2019PASP..131l4504M} coadds.
unWISE incorporates additional data from the reactivation of the satellite as NEOWISE, giving $\approx$0.7 magnitudes deeper coverage in \textit{W1} and \textit{W2} compared to AllWISE.
The \citet{2019PASP..131l4504M} coadds are also deeper than the \citet{2014AJ....147..108L} images which were force-photometered to produce the \textit{WISE} measurements reported in the \citet{DR16Q} catalogue.
The unWISE catalogue is matched to SDSS using a 3.0\,arcsec matching radius, keeping only sources with unique matches within that radius. unWISE models the \textit{W1} and \textit{W2} pixel data separately, and to reduce the number of contaminants we keep only sources where there is a detection in both \textit{W1} and \textit{W2}. 
We find  650\,551 unWISE sources match to quasars from SDSS DR16.
As shown in Fig.~\ref{fig:completeness}, this includes more than 98 per cent of all quasars with $i_\textrm{AB}<20.1$.

\subsubsection{UKIDSS photometry}

We cross-match the SDSS DR16 quasar catalogue to the eleventh data release of the UKIDSS Large Area Survey \citep[UKIDSS-LAS DR11plus;][]{Lawrence07}.
UKIDSS-LAS covers almost three tenths of the SDSS footprint in the near-infrared \textit{YJHK} bands
\citep{2006MNRAS.367..454H, 2009MNRAS.394..675H}.
Compared to previous releases of UKIDSS data, DR11plus includes a slight increase in the fraction of the SDSS footprint which is covered, and a significant improvement to the \textit{J}-band depth through the co-addition of a second epoch of imaging.
We use  `apermag3' values, which are the default point source  2.0\,arcsec diameter aperture corrected magnitudes.
We match 152\,946 
% DR14: 116\,370
quasars from SDSS DR16 to unique sources within 0.85\,arcsec in the UKIDSS-LAS DR11plus catalogue. This number of matches is slightly larger than the number of force-photometered detections given by \citet{DR16Q}, and we believe the improvement is due to our use of more recent data from UKIDSS.
Of the 650\,551 unWISE-DR16Q matches, 139\,396 quasars are also matched to UKIDSS-LAS DR11plus.
However, the  limiting magnitude of UKIDSS is brighter than that of SDSS, and so the completeness of \textit{YJHK} matching drops significantly above $i_\textrm{AB} \approx 19.1$.

\subsubsection{Galactic extinction} 

We correct the \textit{ugrizYJHK} photometric bands for Galactic extinction using the dust maps of \citet{Schlafly10} and \citet{Schlafly11}. 
Such extinctions are typically of the order $E(B-V)<0.1$, 
as most SDSS quasars lie outside the Galactic plane.
%We remove 63 objects from the DR16 catalogue with Galactic extinction $E(B-V)>0.3$.
%These 63 objects all lie outside the UKIDSS-LAS footprint

Commonly quoted passband attenuations are correct only when the source SED is similar to that of an elliptical galaxy at low redshift. 
A type 1 quasar SED is bluer than the SED of a typical star or galaxy, leading to subtle differences in the conversion from $E(B-V)$ to the attenuation in each observed passband.
We  derive our own passband attenuations using a $z=2.0$ quasar source SED,
assuming $R_V=3.1$ and the Galactic extinction curve of \citet{Fitzpatrick09}. The conversions we use are given in Table~\ref{tab:galext}.  
Using source quasar SEDs in the range $0\le z \le 5$ alters these values by no more than three per cent.
The attenuation due to dust in the \textit{WISE} bands is negligible for the extinctions we consider and no correction for Galactic extinction is applied to these data.

\begin{table}
\centering
\caption{\small Passband attenuations $A_\lambda/E(B-V)$ for \textit{ugrizYJHK} filters adopting the $R_V=3.1$ Galactic reddening law from \citet{Fitzpatrick09} and a $z=2.0$ quasar SED.}
\label{tab:galext}
\begin{tabular}{l |c }
%\hline
Filter & $A_\lambda/E(B-V)$ \\
\hline
SDSS \textit{u} & 4.82\\
SDSS \textit{g} & 3.80\\
SDSS \textit{r} & 2.58\\
SDSS \textit{i} & 1.92\\
SDSS \textit{z} & 1.42\\
UKIDSS \textit{Y} & 1.12\\
UKIDSS \textit{J} & 0.80\\
UKIDSS \textit{H} & 0.52\\
UKIDSS \textit{K} & 0.33\\
\hline
\end{tabular}

\end{table}

\subsection{Sample definition and binning}
\label{sec:binning}

As we expect the average quasar colour at any given redshift to change as a function of luminosity,
we calibrate our model parameters using the  colours of quasars which have been binned in apparent magnitude (i.e., flux) as well as in redshift. The numbers of objects contributing to each bin are given in Table~\ref{tab:data}.

\begin{table}
 \centering
\caption{\small
The number of quasars  remaining at each stage of  the cross-matching of catalogues described in Section~\ref{sec:photometry}, and in  construction of the three flux bins used to calibrate our model parameters in Section~\ref{sec:binning}.
}
\begin{tabular}{ l | r}

  & No. of quasars\\
\hline
 SDSS DR16 quasar catalogue, $0<z<5$ & 748\,620 \\
\hline
  Matched to unWISE  (\textit{ugrizW12}) & 650\,551 \\
Brighter flux bin:\\ 
 \hspace{1cm} $17.6<i_\textrm{AB}<18.1$ & 9903 \\
 \hspace{1cm} 107 redshift bins & $\simeq$93 per bin \\
Fainter flux bin:\\
 \hspace{1cm} $19.6<i_\textrm{AB}<20.1$ & 103\,290 \\
 \hspace{1cm} 107 redshift bins & $\simeq$965 per bin \\
\hline
  Matched to UKIDSS-LAS  (\textit{ugrizYJHKW12}) & 139\,396\\
Fiducial flux bin:\\
 \hspace{1cm} $18.6<i_\textrm{AB}<19.1$ & 15\,203 \\
 \hspace{1cm} 214 redshift bins & $\simeq$71 per bin \\
\hline
\end{tabular}
\label{tab:data}
\end{table}

First, we take all objects with SDSS-UKIDSS-unWISE \textit{ugrizYJHKW12} photometry in the flux range $18.6<i_\textrm{AB}<19.1$.
The completeness of SDSS quasars within the UKIDSS-LAS drops significantly above $i_\textrm{AB}=19.1$, and the number of objects with $i_\textrm{AB}<18.6$ is relatively low, and so 
we then take all objects with SDSS-unWISE \textit{ugrizW12} photometry in the flux ranges $17.6<i_\textrm{AB}<18.1$ and $19.6<i_\textrm{AB}<20.1$ (i.e. one magnitude brighter and one magnitude fainter than the UKIDSS-matched flux bin). 

Within each flux range, redshift bins are spaced such that there are equal numbers of objects in each redshift bin.
Our fiducial $18.6<i_\textrm{AB}<19.1$ flux range is divided into 214 redshift bins, and our brighter and fainter flux ranges are each divided into 107 redshift bins.

Within each flux-redshift bin, the distribution of each colour is roughly Gaussian, with  a  tail to redder colours  which we ascribe to dust at the quasar redshift.
This tail is found to be stronger in the fainter flux bins, which is consistent with the fact that reddening is a side-effect of wavelength-dependant extinction, i.e. the flux in reddened objects is attenuated and so they also appear fainter.
We therefore apply an iterative sigma-clipping procedure to exclude this tail, and any other outliers, when calculating the median colour in any given flux-redshift bin. We clip objects more than 2 sigma away from the sample median, and use the standard deviation of the resulting clipped sample as a measure of the dispersion around the median.

\subsection{Fitting routine}

For every colour in each of the flux-redshift bins described above, we compute the median $M_{z,f}$ and dispersion $\sigma_{z,f}$ of the sigma-clipped colour distribution.
We also compute the median redshift $z$ and median absolute magnitude $M_i$ in each bin.
For a given set of  model parameters, the SED model is evaluated at
these values of $z$ and $M_i$, giving a set of synthetic colours $S_{z,f}$.
The loglikelihood is then defined in the usual way as
\begin{equation}
    \ell = -\frac{1}{2} \sum_{z,f}\left(\frac{(M_{z,f} - S_{z,f})^2}{\sigma_{z,f}^2}
           + \textrm{log}({\sigma_{z,f}^2})\right)
\label{eq:likelihood}
\end{equation}
where we exclude the colours which, for a given redshift, probe
rest wavelengths $\lambda<912$\,\AA\ or $\lambda>3$\,\micron.
In practice this means we exclude $W1-W2$ at $z<0.7$, $K-W1$ at $z<0.3$, $g-r$ at $z>3.0$ and $u-g$ at $z>2.3$.
Using flat priors on all parameters, samples are drawn from the posterior distribution using an affine-invariant ensemble Markov Chain Monte Carlo method \citep{2010CAMCS...5...65G}.
We use 200 walkers with 750 steps each, which are started near the least-squares solution to Eq.~\ref{eq:likelihood}. We remove the first 250 steps from each walker, having verified that this removes the burn-in and the chains are subsequently well-mixed.

\subsection{Results}

\begin{figure}
    \centering
    \includegraphics[width=\columnwidth]{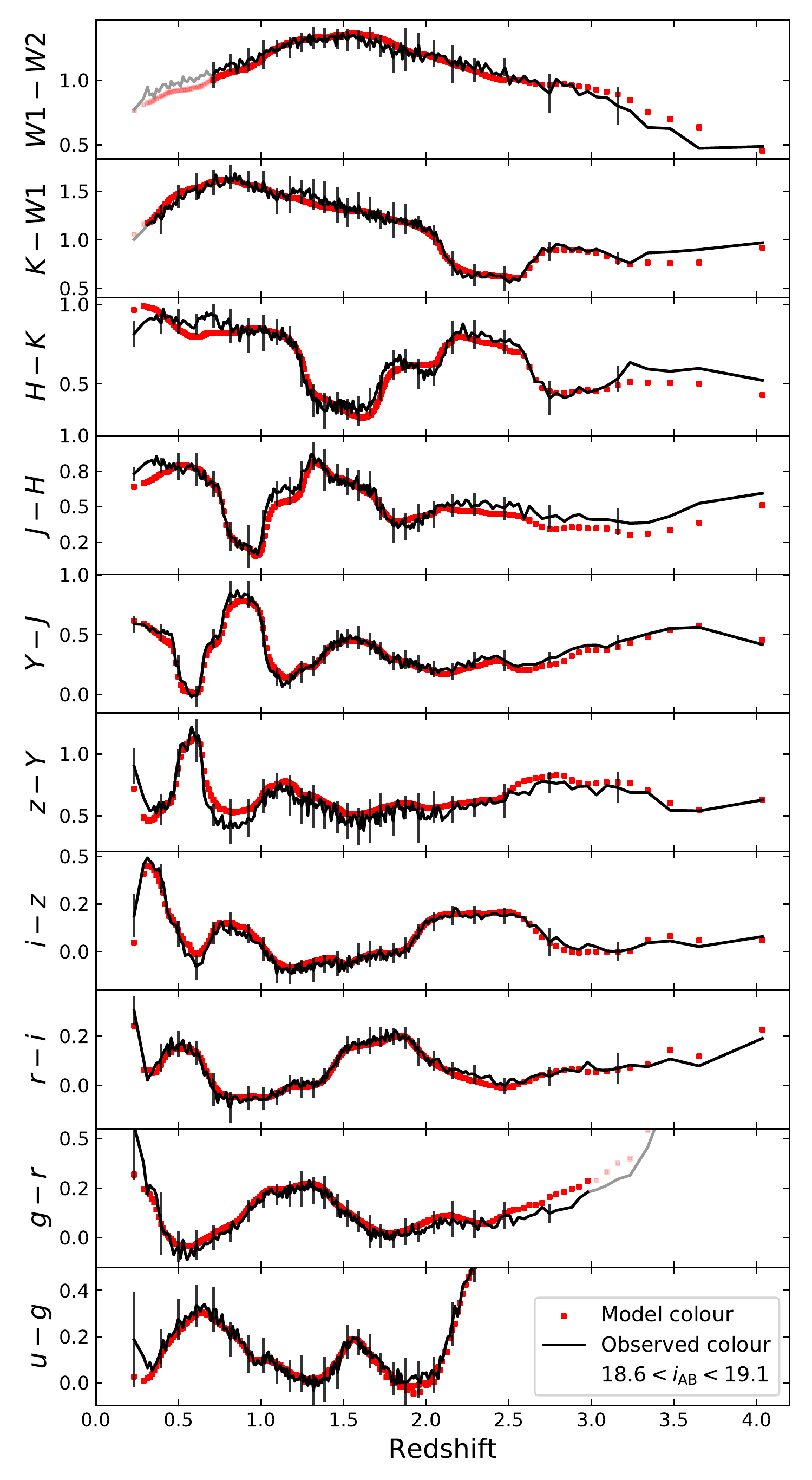}
    \caption
    {Black:  median observed SDSS-UKIDSS-unWISE quasar colours in each redshift bin for our sample of  $18.6<i_\textrm{AB}<19.1$ quasars. Error bars show the dispersion around the median in every eighth bin. Red: the  calibrated model. Rest wavelengths $\lambda<912$\,\AA\ or $\lambda>3$\,\micron, where the model is not fit to the data, are grayed out.}
    \label{fig:colour_redshift_tracks}
\end{figure}

In Fig.~\ref{fig:colour_redshift_tracks} we show the median sigma-clipped \textit{ugrizYJHKW12} colours in our $18.6<i_\textrm{AB}<19.1$ flux bin. The number density of objects is such that there are more redshift bins in the range $1<z<2$ than there are in $3<z<5$. Significant deviations in colour are seen as a function of redshift, due to strong emission lines such as Ly$\alpha$ and H$\alpha$ moving in and out of the different photometric filters. Error bars show the dispersion within every eighth bin.

\begin{figure}
    \centering
    \includegraphics[width=\columnwidth]{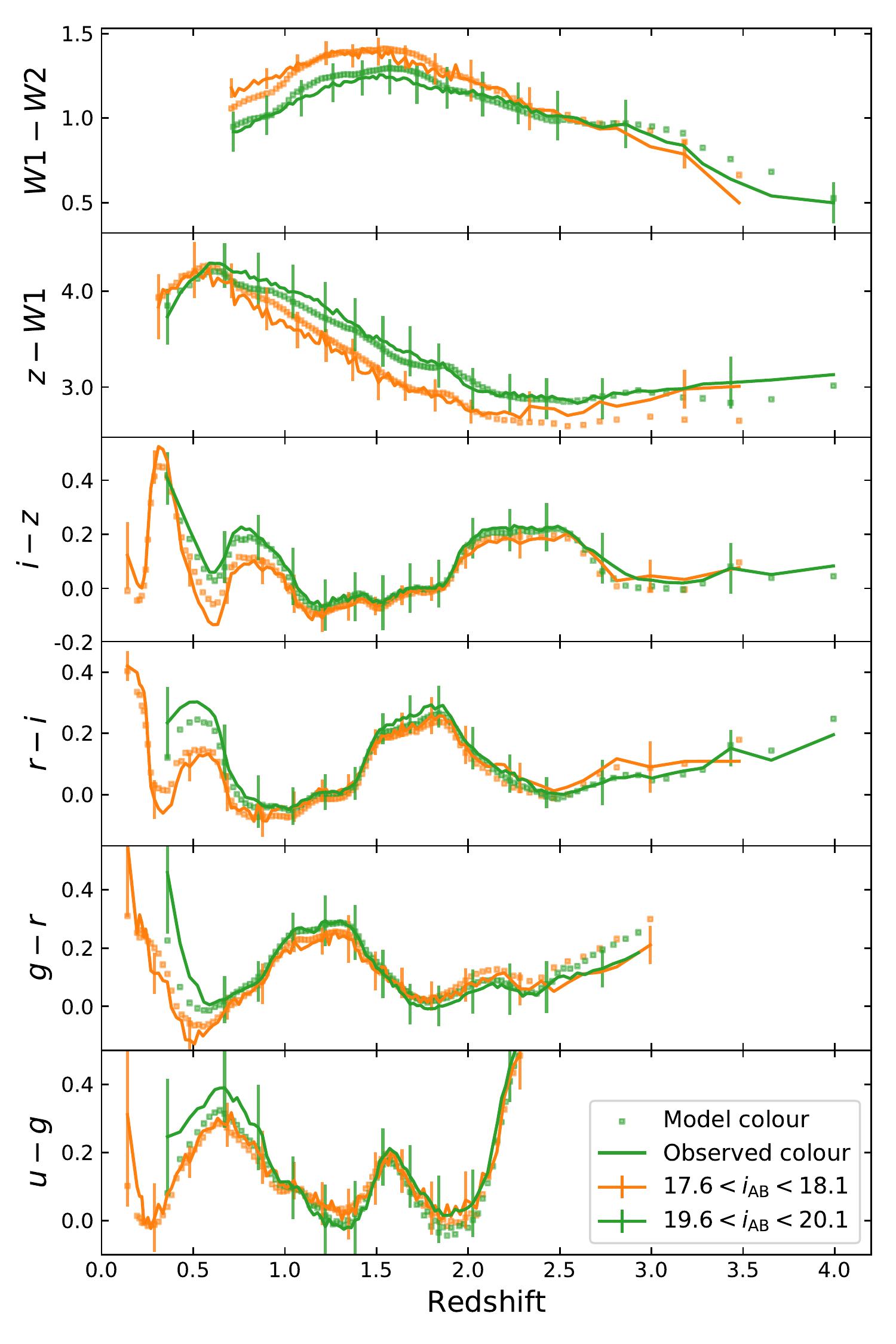}
    \caption
    {As Fig.~\ref{fig:colour_redshift_tracks}, but showing the SDSS-unWISE quasar colours in $17.6<i_\textrm{AB}<18.1$ (orange) and $19.6<i_\textrm{AB}<20.1$ (green), together with the calibrated model for each apparent magnitude range (squares).
    The colours of fainter quasars have a larger contribution, due to starlight from the host galaxy, to the quasar SED, which accounts for the variation in \textit{W1} and \textit{W2} and the low-redshift changes in \textit{ugriz}.
    Smaller differences in \textit{ugriz} at redshifts $z>1$ are due to changes in the equivalent width of the emission lines, which we model as a Baldwin effect using the templates in Fig.~\ref{fig:emlines}. }
    \label{fig:colours_imagbins}
\end{figure}

In Fig.~\ref{fig:colours_imagbins} we  show the SDSS-unWISE \textit{ugrizW12} colours for our brighter and fainter flux bins.
Changes in the average quasar colours (at fixed redshift) are expected with varying luminosity, due to a varying fraction of host galaxy flux contribution.
This is expected to be most significant around the 1\,\micron\ minimum in the quasar SED, where the optical continuum is falling but the hot dust emission is yet to start rising.
At redshifts $1<z<2$, this minimum is moving through the unWISE \textit{W1} passband, and in Fig.~\ref{fig:colours_imagbins} we can see that there are noticeable changes in the \textit{zW1W2}
colours between our bright and faint flux bins.
At rest-frame optical wavelengths, the flux due to starlight is generally going to be redder than the quasar continuum, so a larger host galaxy contribution has the effect of reddening the \textit{ugriz} colours at lower redshifts $z < 1$.
Finally, the equivalent width of the ultraviolet emission lines is expected to increase as the luminosity of the quasars decreases, the so-called Baldwin effect, leading to slightly larger-amplitude wiggles in the \textit{ugriz} colour-redshift tracks at higher redshifts $z>1$.

In Figs.~\ref{fig:colour_redshift_tracks} and \ref{fig:colours_imagbins}, we also show predicted colours from our calibrated model.
The maximum likelihood value for each model parameter is given in Table~\ref{tab:SEDparams},
and the full posterior distribution is given in Appendix~\ref{sec:corner}.
The normalisation of the Balmer Continuum is found to be consistent with zero, which is to be expected given that the emission-line templates we use have not had any Balmer Continuum subtracted.
The model is seen to be very good at reproducing the average quasar properties across the vast majority of the colour-redshift-flux space.


\begin{table}
\centering
    \caption{Free parameters in our quasar SED model (Section~\ref{sec:model}) and their values after calibrating to SDSS, UKIDSS and unWISE data in Section~\ref{sec:params}.}
    \label{tab:SEDparams}
    \begin{tabular}{ r r c}
        Description & Parameter & Value \\
        \hline
        blue power-law slope $\alpha_1$ & \texttt{plslp1} & -0.349 \\
        red power-law slope $\alpha_2$ & \texttt{plslp2} & +0.593 \\
        power-law break wavelength $\lambda_\textrm{break}$ [\AA] & \texttt{plbrk1} & 3880 \\
        blackbody normalisation & \texttt{bbnorm} & 3.96\\
        blackbody temperature $T_\textrm{BB}$ [K] & \texttt{tbb} & 1240 \\
        overall emission line scaling & \texttt{scal\_em} & -0.994 \\
%        \ha\ emission line scaling & \texttt{scal\_ha} & 1.0 \\
%        \lya\ emission line scaling & \texttt{scal\_lya} & 1.0 \\
%        Narrow emission line scaling & \texttt{scal\_nlr} & 1.0 \\
        Baldwin Effect `slope' & \texttt{beslope} & 0.183 \\
        galaxy fraction at $M_i=-23$ & \texttt{fragal} & 0.244 \\
        galaxy luminosity power-law index & \texttt{gplind}  & 0.684 \\
        \hline
    \end{tabular}
\end{table}

We have successfully constructed a model which reproduces the median SDSS-UKIDSS-unWISE colours of a sample of quasars at $17.6<i_\textrm{AB}<20.1$. However, future surveys such as LSST will probe significantly fainter than this flux limit, so we now need to verify that our model produces accurate predictions at fainter apparent magnitudes.

\section{Comparison with other observations}
\label{sec:verif}

We have constructed a model which is capable of reproducing the colours of relatively bright ($i_\textrm{AB}<20.1$) quasars with redshifts $0<z<5$ in SDSS, UKIDSS and unWISE. 
Our model includes two components which adjust the resulting colours as a function of luminosity: both the normalisation of the galaxy component and type of emission-line template depend on the quasar luminosity.

Before providing predictions for upcoming surveys in Section~\ref{sec:predict}, here we verify that our model matches observations of known quasar populations in regions of redshift-luminosity space which have not been used to inform the construction of the model.
We compare our predicted model colours to observations of two populations of quasars which have not been used to constrain our model parameters in any way: 
the near-infrared colours of quasars which are 2.4 magnitudes fainter than the UKIDSS-matched sample used in Section~\ref{sec:params}, and known $z>5$ quasars.

\subsection{Fainter quasars}

We take VISTA \textit{ZYJH} data \citep{2018MNRAS.474.5459G} from the fifth data release of the ESO public survey VIKING  \citep{Edge13}. VIKING DR5 is cross-matched to the SDSS DR16 quasar catalogue using a 0.85\,arcsec matching radius, yielding a total of 17\,047 objects.

\begin{figure}
    \centering
    \includegraphics[width=\columnwidth]{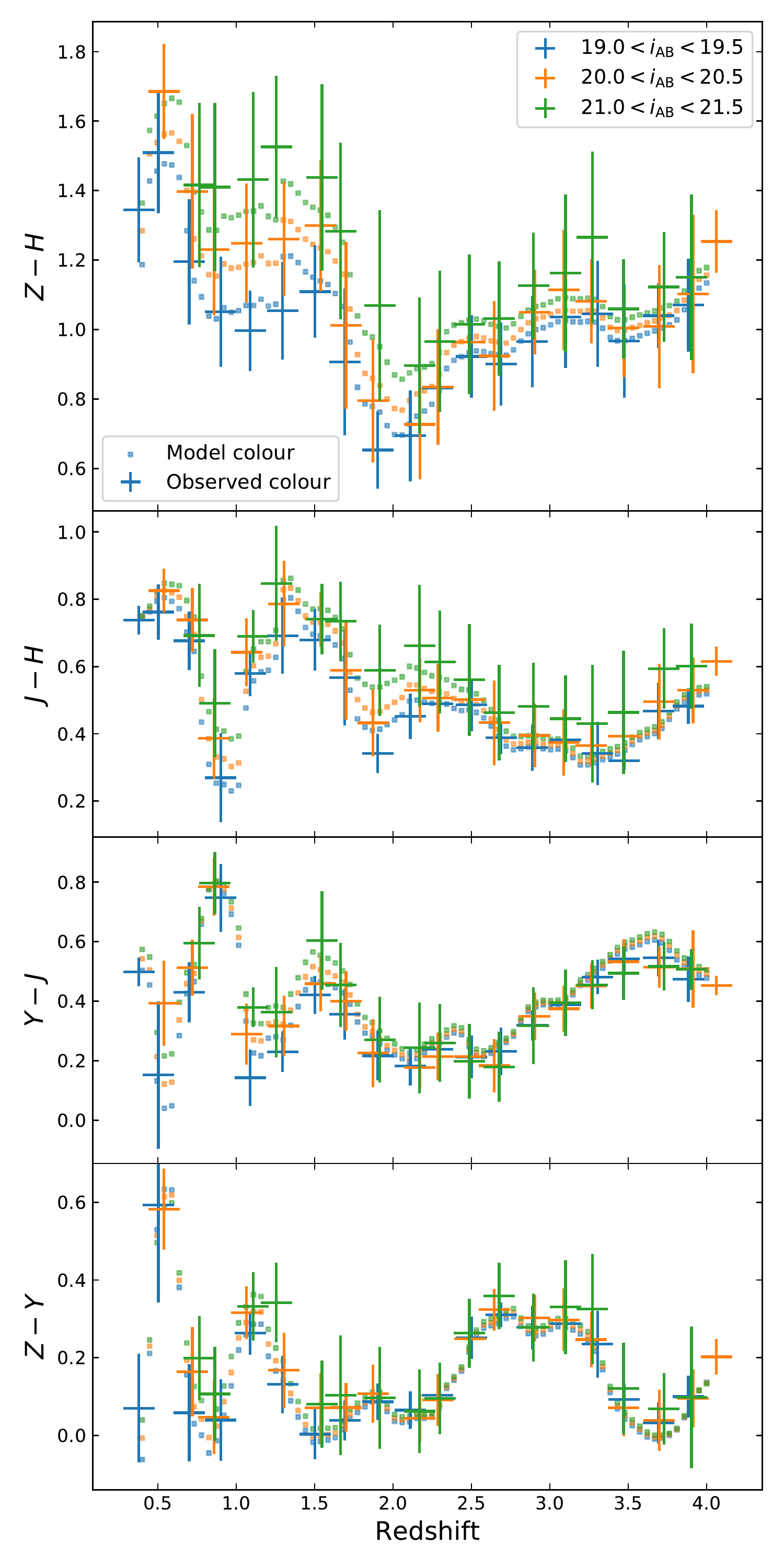}
    \caption{VIKING \textit{ZYJH} colours in three flux bins, probing 2.4 magnitudes fainter than UKIDSS data used to constrain our model. Error bars show the observed average colours and standard deviation within each redshift bin, and squares show the predicted model colours. Our model predictions  reproduce the changes in colour which are observed as a function of  luminosity, giving us confidence in our ability to extrapolate to LSST-like depths.}
    \label{fig:VIK_colours}
\end{figure}

In Fig.~\ref{fig:VIK_colours}, we show the median sigma-clipped VIKING colours in redshift bins of width $\Delta z=0.2$. Only bins containing 10 or more objects are shown. The observed colours are seen to change significantly as a function of apparent magnitude, with the \textit{Z-H} colours at $i_\textrm{AB}\simeq21$ up to 0.4\,mag redder than those at  $i_\textrm{AB}\simeq19$. 
Our model is seen to reproduce the observed median colours to within $\simeq$0.1\,mag, significantly less than the observed intra-sample dispersion, across all redshifts and across all three flux bins.

\subsection{High redshift quasars}

\begin{figure}
    \centering
    \includegraphics[width=\columnwidth]{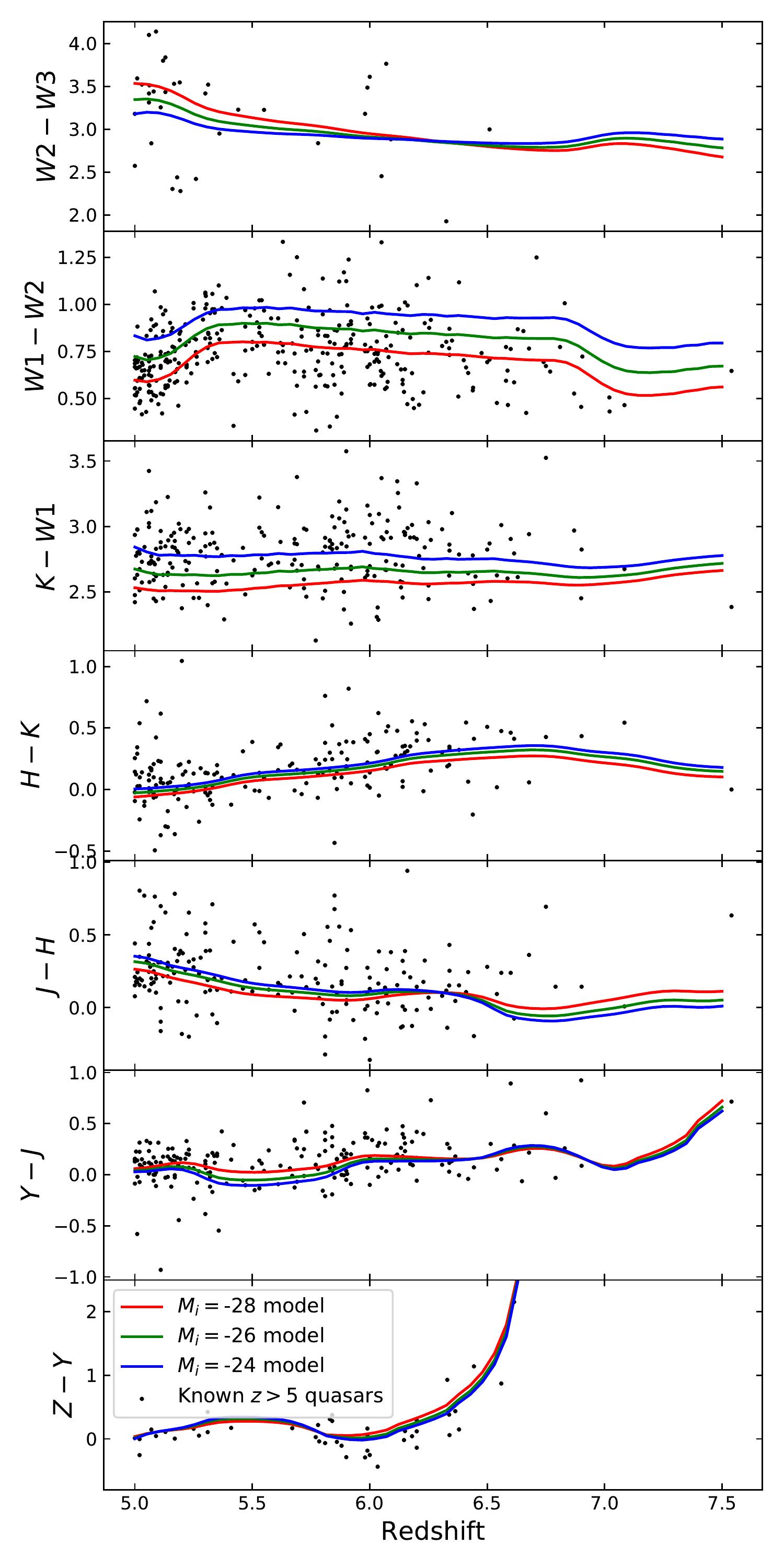}
    \caption{Black points: the observed colours of individual $z>5$ quasars from \citet{2020MNRAS.494..789R}. Any colours for a quasar involving a magnitude uncertainty $>$0.3\,mag are excluded.
    Note that the errors on these colours are highly asymmetric: for example \textit{W1} is significantly deeper than \textit{K} and so the \textit{K-W1} colours  preferentially scatter to larger values.
    Coloured lines: our predicted model colours for a range of intrinsic luminosities, corresponding to a range of host galaxy contributions, assuming $E(B-V)=0$.}
    \label{fig:Hz_colours}
\end{figure}

In Fig.~\ref{fig:Hz_colours},
we show the observed colours of known $z>5$ quasars using force-photometered data from \citet{2020MNRAS.494..789R}. 
We remove all photometric measurements which have  uncertainties greater than 0.3 mag.
The median uncertainties in each band for the remaining data are 
$[\sigma(Z), \sigma(Y), \sigma(J), \sigma(H), \sigma(K), \sigma(W1), \sigma(W2), \sigma(W3)] = [0.05,  0.07,  0.08, 0.10, 0.09, 0.04, 0.07, 0.23]$
respectively.
The errors  on the derived colours are therefore asymmetric: for example the uncertainty of \textit{W1} is significantly less than that of \textit{K} and so the \textit{K-W1} colours preferentially scatter to larger values.
We also show our predicted model for three different apparent magnitudes, chosen to be representative of the luminosity range of the $1<z<5$ SDSS quasar population. The large changes in predicted colours seen in \textit{K-W1} and  \textit{W1-W2} reflect the variation in host galaxy contribution  in our model with changing luminosity, while the changes in \textit{Z-Y} and \textit{Y-J} at these redshifts reflect changes in the average emission-line properties.
Many of the quasars in the \citet{2020MNRAS.494..789R} compilation have errors in the colours $>$0.2\,mag contributing to the extended spread of colours at fixed redshift. Overall, the model predictions are consistent with the observed colours within the uncertainties on the data.

\section{Predicted Quasar colours for LSST and Euclid}
\label{sec:predict}

\begin{figure}
    \centering
    \includegraphics[width=\columnwidth]{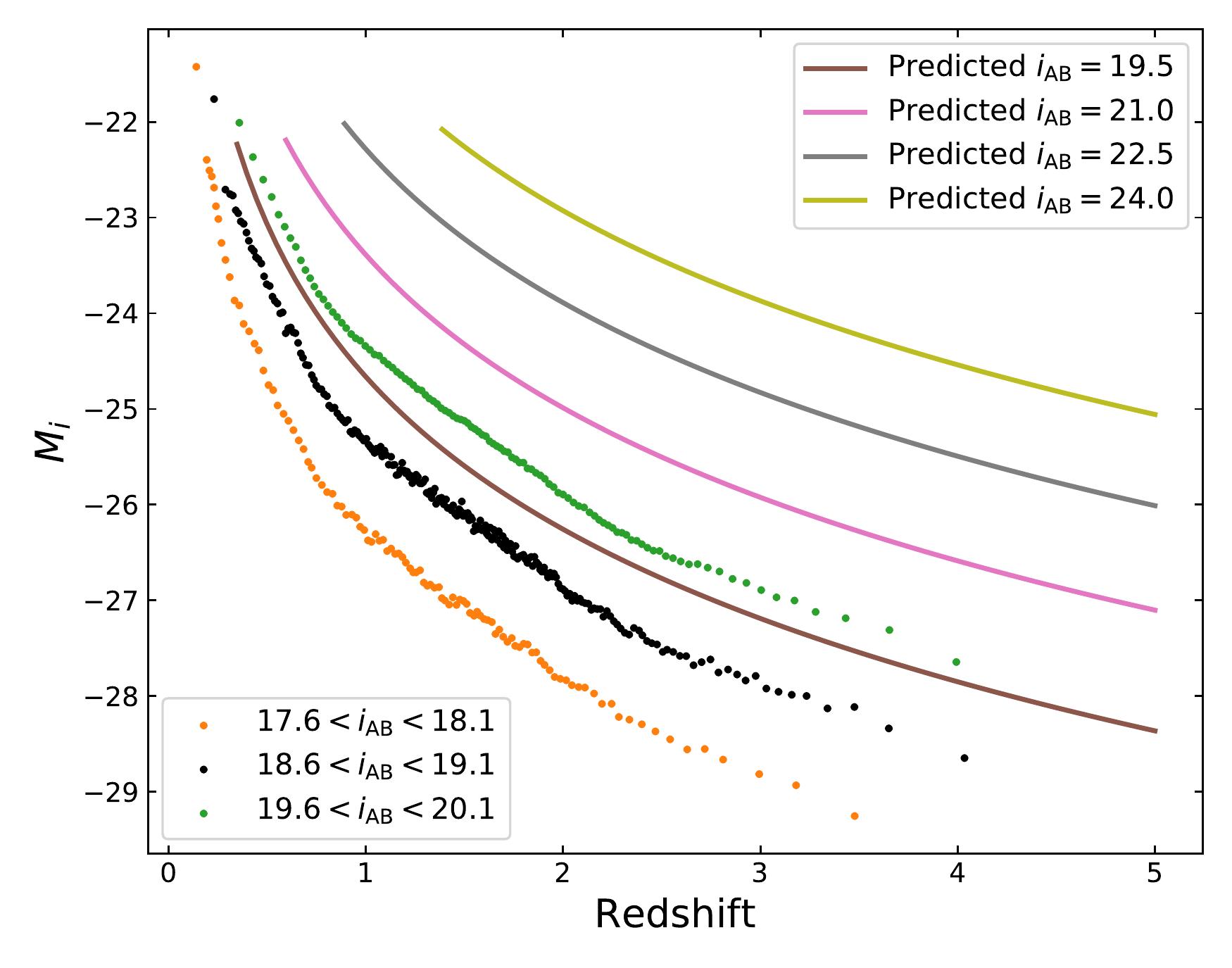}
    \caption{
    Coloured points: the median $M_i$ in each flux-redshift bin presented in Fig.~\ref{fig:colour_redshift_tracks} and Fig.~\ref{fig:colours_imagbins}.
    Solid lines: predicted luminosity-redshift relations (Eq.~\ref{eq:zlum}) assumed when predicting colours for different flux-limited samples at LSST depths (Fig.~\ref{fig:predicted_colours}). We limit our predictions to redshifts and apparent magnitudes where $M_i$ is predicted to be less than $-22.0$, which has the effect of limiting our predictions at the faintest flux limits to higher redshifts.}
    \label{fig:zlum_lumval}
\end{figure}

In order to compute model colours at fainter flux limits, such as those expected to be reached  by LSST, it is first necessary to predict the redshift-luminosity relation used to compute the galaxy normalisation (Eq.~\ref{eq:galnorm}). 
In Fig.~\ref{fig:zlum_lumval}, we show the median observed $M_i$ as a function of redshift for each \textit{i}-band--limited sample. These are found to be approximated by a function of the form
\begin{equation}
    M_i =  -\left[0.25\left(\frac{i_\textrm{AB}}{20}\right) + 5.05\right]\textrm{log}_{10}(z)
    - \left[17.4\left(\frac{20}{i_\textrm{AB}}\right) + 6.82\right].
    \label{eq:zlum}
\end{equation}

The faintest object from SDSS which we have used in calibrating our model is at approximately $M_i=-22.0$, and  we chose not to extrapolate any fainter than this luminosity. In effect this limits our model predictions to redshifts $z>1.5$ for  $i_\textrm{AB}=24.0$  and to redshifts $z>1.0$ for $i_\textrm{AB}=22.5$.

\begin{figure}
    \centering
    \includegraphics[width=\columnwidth]{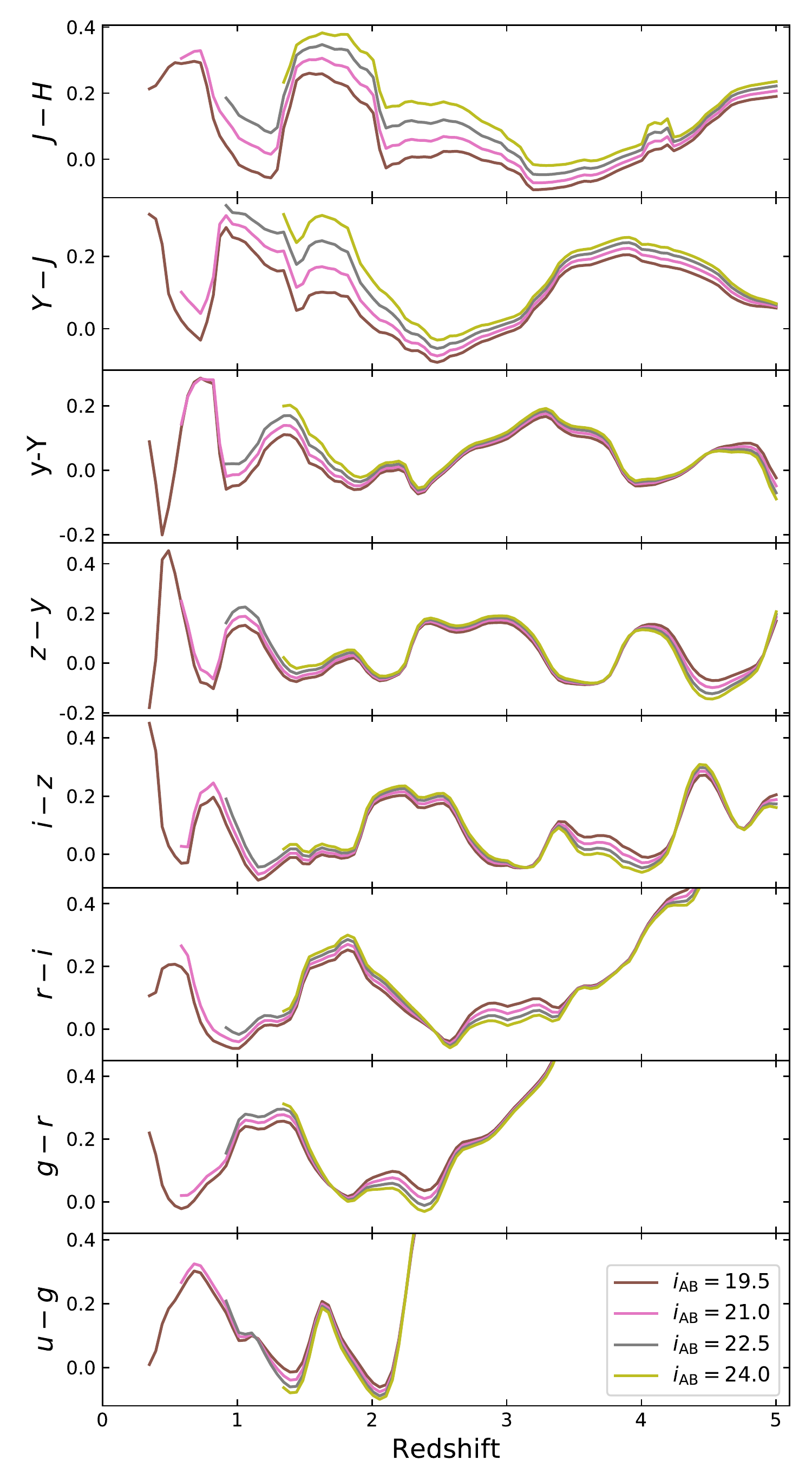}
    \caption{Predicted average colours (assuming $E(B-V)=0$) as a function of redshift for LSST \textit{ugrizy} and \textit{Euclid} \textit{YJH}. Colours are presented on the AB system, and shown for different apparent magnitudes. The host galaxy contribution is predicted to have a noticeable  effect on the \textit{YJH} colours, even at $z>2$, as we move towards the LSST single-epoch limiting magnitude of $i_\textrm{AB}=24.0$.}
    \label{fig:predicted_colours}
\end{figure}
\begin{figure}
    \centering
    \includegraphics[width=\columnwidth]{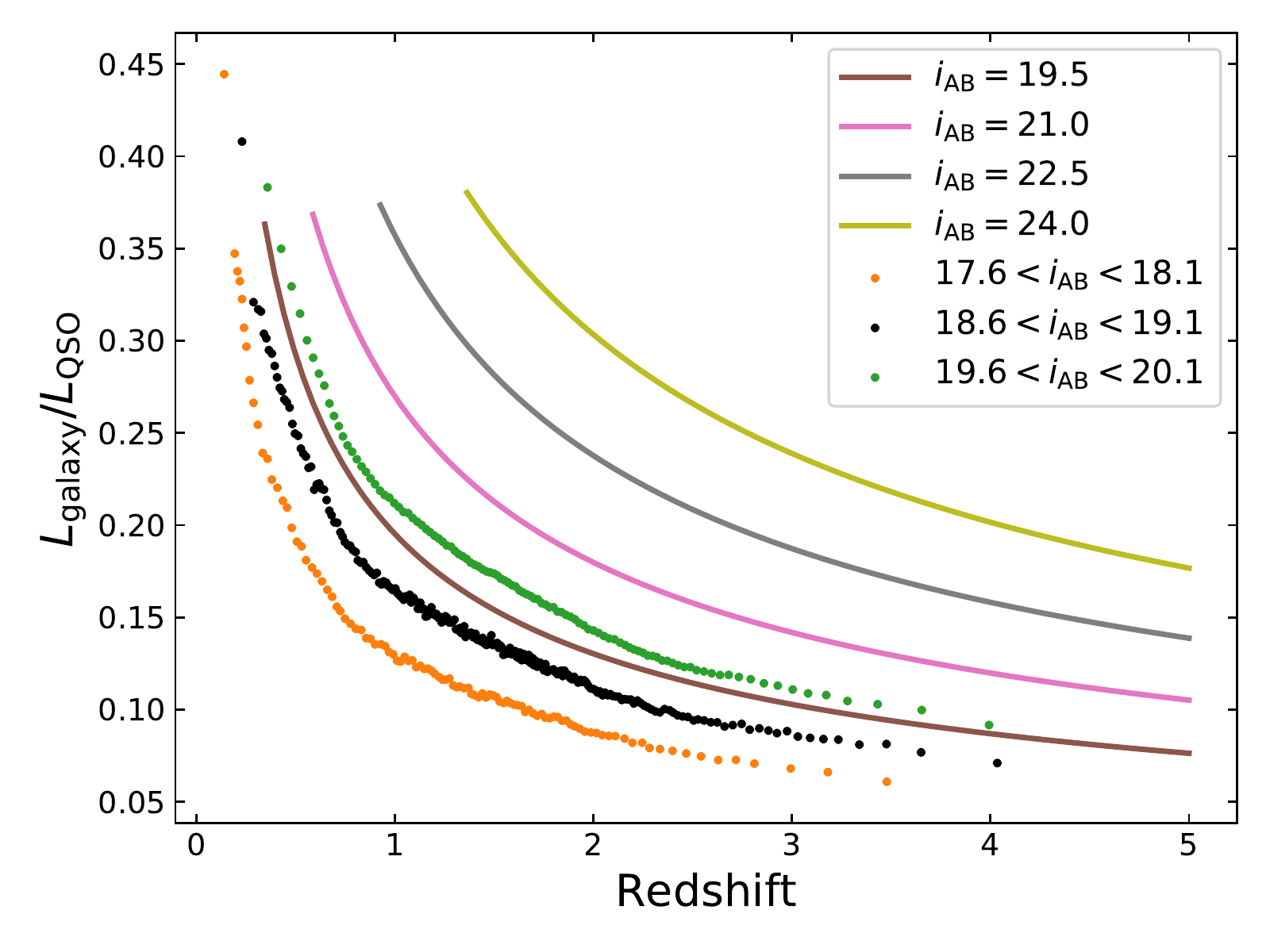}
    \caption{Ratio of integrated 4000-5000\,\AA\ luminosities for the galaxy and quasar components of our SED model (Eq.~\ref{eq:galnorm}).
    Points: the galaxy fraction in our model for the three observed flux regimes.
    Solid lines: predicted models from extrapolating  those trends to LSST-like depths. }
    %The limit of $M_i<-22.0$ roughly corresponds to $L > 10^{44.8}$\ergps.}
    \label{fig:predicted_galfrac}
\end{figure}

Using our predicted $z-M_i$ tracks, we can make predictions for the average colours of quasars in LSST and \textit{Euclid} down to $i_\textrm{AB}=24$, which is the single-epoch limiting magnitude for LSST. 
These are shown in Fig.~\ref{fig:predicted_colours},
and the predicted flux contribution from the host galaxy to the median colours is shown in Fig.~\ref{fig:predicted_galfrac}.
The strong dependence of the fractional galaxy contribution on apparent magnitude leads to significant changes in the predicted \textit{Euclid} \textit{YJH} colours at  redshifts $1<z<3$, with fainter objects having a stronger galaxy contribution and hence redder near-infrared colours.
Here we have assumed that the average host galaxy SED is the same for SDSS quasars as it will be for the fainter \textit{Euclid} population. There are many reasons why this might not be true, and so
it is important not to take our predicted quasar colours as inferring anything about the physics of fainter AGN, but instead merely providing testable predictions for how quasar selection techniques may need to adapt to identify objects in new regimes of luminosity-redshift space.

\subsection{Very-high redshift quasars in Euclid}

\begin{figure}
    \centering
    \includegraphics[width=\columnwidth]{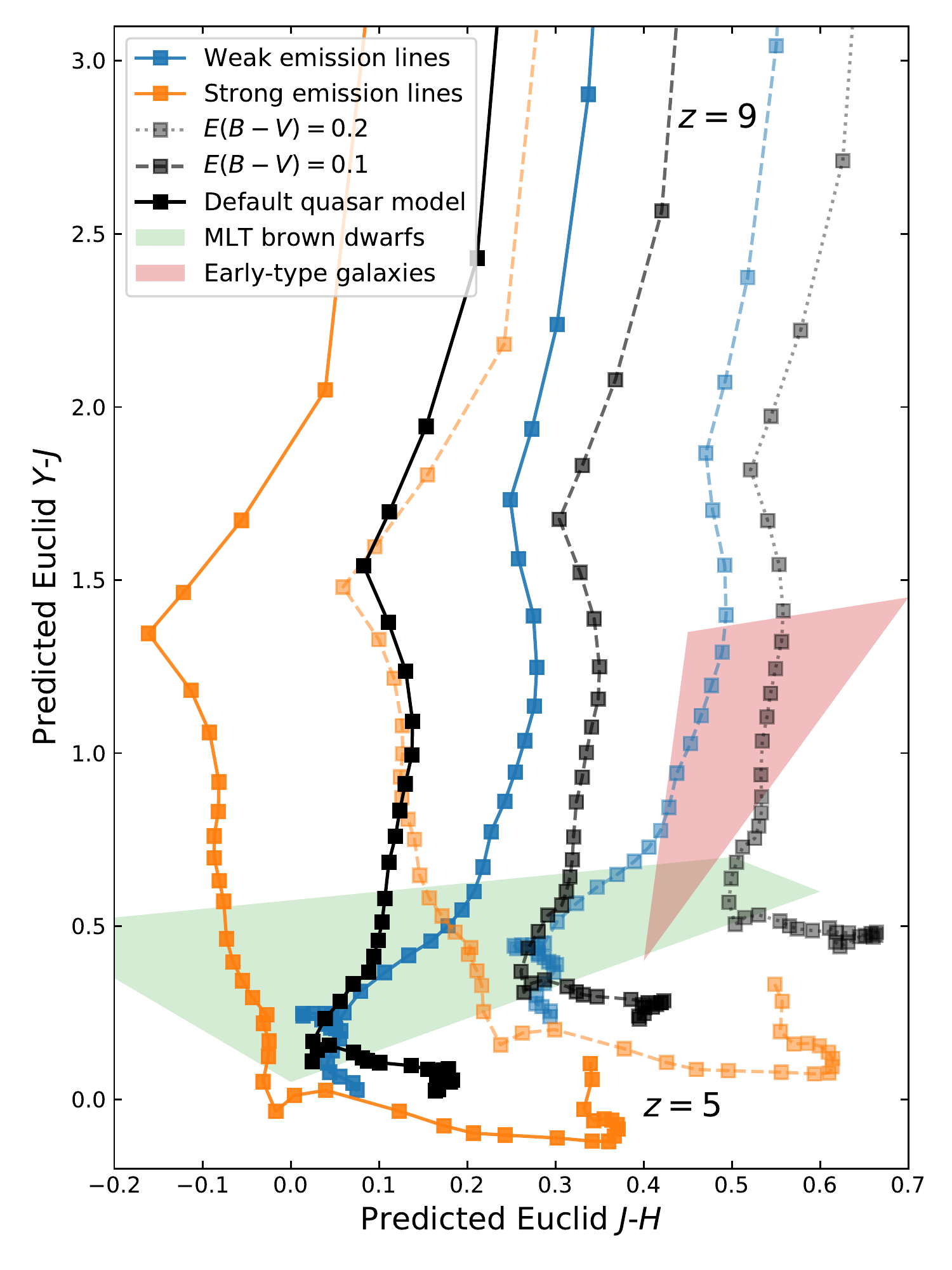}
    \caption{Predicted \textit{Euclid} \textit{YJH} colour-colour tracks for $5<z<9$ quasars,
    %. Colours are presented on the AB system, 
    with each square representing a redshift interval $\Delta z=0.1$. Orange, black and blue tracks show the effect of different emission line templates, and the dashed and dotted tracks show the effects of increasing dust reddening.
    Luminous, high $L/L_\textrm{Edd}$ quasars are expected to have relatively weak emission lines, and so have broad-band colours somewhere between the black and blue tracks.
    The colours of high-redshift quasars in the discovery space opened by \textit{Euclid} could span several tenths of magnitude in \textit{J-H}, which should be taken into account when searching for $z>7$ quasars.
    For reference, we shade regions where relevant contaminant populations of brown dwarfs and early-type elliptical galaxies are expected to be found
    \citep[][fig. 1]{2019A&A...631A..85E}.
    }
    \label{fig:Euclid_colours}
\end{figure}

In Fig.~\ref{fig:Euclid_colours}, we show the \textit{Euclid} \textit{YJH} colour-colour tracks for our quasar model SED with redshifts $z>5$. Square symbols are plotted at redshift intervals of $\Delta z =0.1$.
Varying the emission-line properties of our model between the extrema seen in lower-redshift quasars produces changes in the \textit{Euclid} colours of up to a few tenths of a magnitude. Quasars with weak, blueshifted lines (as expected for the most luminous, highly accreting sources), and modest dust reddening ($E(B-V)=0.1$) are predicted to have \textit{J-H} $\simeq0.5$ at $7<z<8$, overlapping the colour tracks of elliptical galaxies of redshift $z\simeq1$ \citep[cf.][fig. 1]{2019A&A...631A..85E}.
The more common elliptical galaxies may impact on the efficient selection of some $z>7$ quasars, reducing the completeness of the census of black hole growth in the early universe.

\section{Discussion}
\label{sec:discuss}

\begin{figure*}
    \centering
    \includegraphics[width=2\columnwidth]{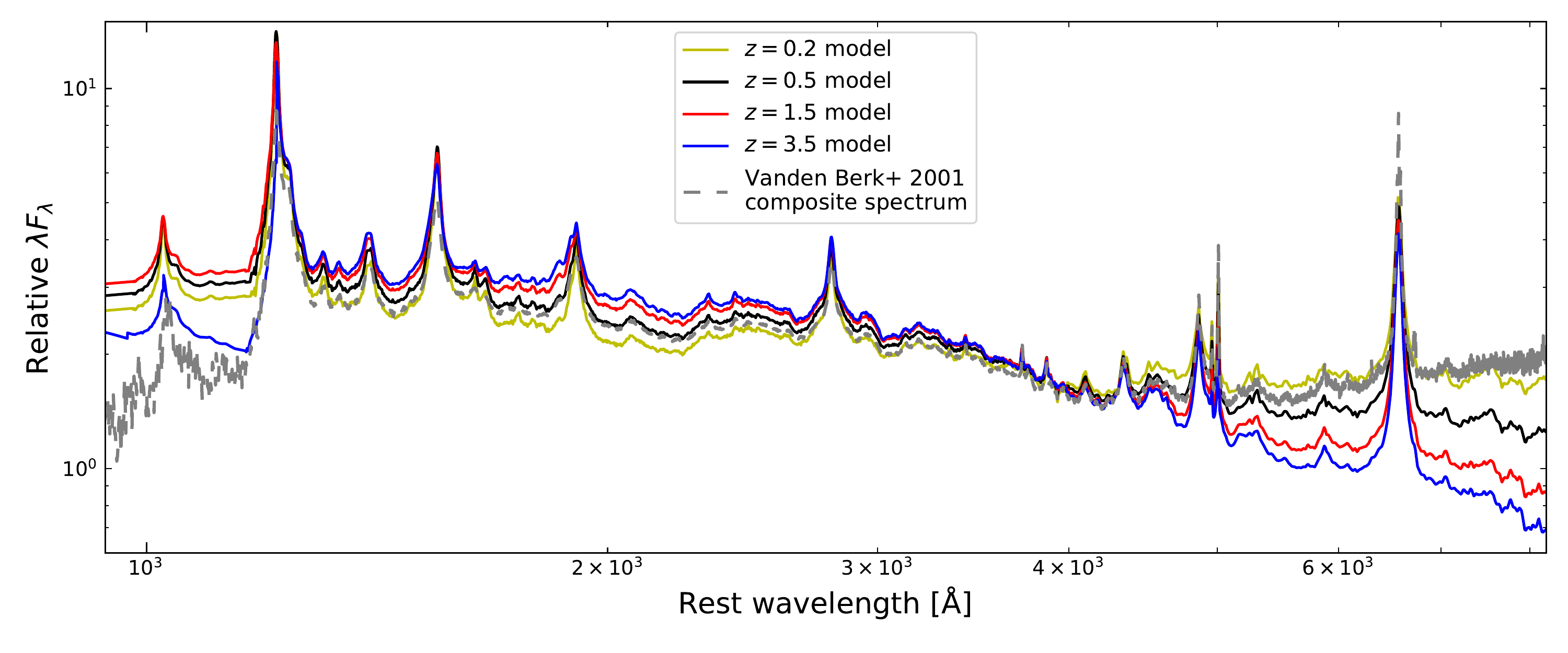}
    \caption{Comparison of our model quasar SED at different redshifts with the composite spectrum from \citet{VandenBerk01}.
    The Vanden Berk composite is reassuringly similar to our models across the rest-frame ultraviolet, and is seen to agree very well with our $z=0.2$ model at rest wavelengths $\lambda>5000$\,\AA, consistent with the redshifts of the objects contributing to the composite at those wavelengths.
    At higher redshifts the average quasar is brighter and has a smaller fractional host galaxy contribution to the rest-frame optical.
    Similarly the composite agrees well with our  $z=3.5$ model at 1200\,\AA, as the inter-galactic medium begins to suppress the flux shortwards of Ly$\alpha$. At $\lambda\approx1000$\,\AA\ the composite consists of objects with $z>4$ where the Lyman suppression is even stronger.
    }
    \label{fig:VandenBerk}
\end{figure*}

A significant improvement over the SED model developed by Hewett is the parametrization of the host-galaxy contribution to the SED as a function of quasar luminosity.
In Fig.~\ref{fig:VandenBerk}, we show the SED model over a range of redshift. Each model has a luminosity chosen at each redshift to match a flux-limited sample similar to SDSS DR7.
At wavelengths longer then 4000\,\AA, the host galaxy contribution increases as one moves to lower redshifts (with fainter AGN), leading to significant changes in the continuum slope in the rest-frame optical region.

In Fig.~\ref{fig:VandenBerk}, we also show the composite of quasar spectra presented by \citet{VandenBerk01}. The Vanden Berk composite made use of observed-frame optical spectra from the SDSS and, as such, the average redshift of objects changes systematically as a function of rest wavelength: pixels at $\lambda<1000$\,\AA\ in the composite are derived from objects with $4<z<5$,
while pixels at $\lambda>5000$\,\AA\ are informed mostly by objects with $z<0.2$.
Reassuringly, and without any fine-tuning, our default model is seen to agree very well with the Vanden Berk composite, when we compare any given wavelength region of the composite with the model for the appropriate redshift. We have also verified that our model agrees with the X-Shooter composite presented by \citet{2016A&A...585A..87S}, which made use of much more luminous ($M_i \approx -29$) quasars at redshifts $1.0 < z < 2.1$.

Our blue power-law slope of $\alpha_\nu=-0.349$ corresponds to $\alpha_\lambda=-1.651$, which is consistent with most composite spectra in the literature: e.g. \citet{2016A&A...585A..87S} find $\alpha_\lambda=-1.70$, \citet{VandenBerk01} find $\alpha_\lambda=-1.56$ bluewards of 5000\,\AA\ and \citet{Francis91} find $\alpha_\lambda=-1.68$. Overall there is close agreement between the ultraviolet continuum and the composites of \citet{Francis91, VandenBerk01, 2002ApJ...565..773T, Glikman2006, 2015MNRAS.449.4204L}. Given the composites span a wide range of luminosities and redshifts, there is no evidence for significant evolution in the shape of the quasar continuum in the rest-frame ultraviolet at these luminosities.

A feature of our SED model, which distinguishes it from some in the literature, is the flexibility afforded by a broken power-law describing the quasar continuum. The second power-law slope describes the shape of the continuum redward of $\approx$4000\,\AA. The value of this red slope is inherently degenerate with host galaxy emission, when the photometric information extends only to the $K$-band. The behaviour of the longer wavelength {\it WISE} \textit{W1-W2} colours as a function of quasar luminosity, however, means it is possible to break the degeneracy between the slope of the red power-law continuum and the contribution from host galaxy. Specifically, the significant range in the \textit{W1-W2} colours as a function of quasar luminosity (at fixed redshift) evident up to $z\approx2$ are well-reproduced by the changing galaxy contribution as a function of quasar luminosity. 

As a result, the dependence of the host-galaxy luminosity on the quasar luminosity, $L_{\rm galaxy} \propto L_{\rm QSO}^{0.684}$, incorporated in the SED model, is more consistent with constraints from other investigations.
\citet{2006ApJS..166..470R} find (their section 5.2) a $L_\textrm{galaxy}-L_\textrm{QSO}$ relation (their eq.~1), which is very similar to our parametrization. Their power-law index, derived from \citet{2006AJ....131...84V}, is 0.87, compared to our 0.684. However, there is also an $L/L_\textrm{Edd}$ term in their eq.~1 which, given brighter quasars will have on average slightly higher accretion rates, will pull their index down closer to ours.

In this work we have assumed a particular functional form (a power-law in luminosity) for the average fractional host galaxy contribution. While this is found to describe the observed data perfectly adequately, we note that there is no \textit{a priori} reason to believe that this is the exact form which the relationship takes.
Working under this assumption, the \textit{WISE} data we use is  of sufficiently high-quality to infer that
the fractional host-galaxy contribution to the rest-frame optical and 1\,$\micron$ flux of luminous quasars is perhaps higher than previously assumed, around 10 to 20 per cent at $z=1$ and 5 to 10 per cent at $z=3$ for SDSS-like flux limits.
We note that the relation found by \citet{2006ApJS..166..470R} represents a \emph{minimal} galaxy contribution and that %\citeauthor{2006ApJS..166..470R}
they themselves admit that it could be much greater.

While there remains a slight degeneracy between the strength of the galaxy emission and the other model parameters  (Fig.~\ref{fig:MCMC_corner}),
it then follows that the value of the second power-law slope, describing the shape of the quasar continuum redwards of 3880 Å, is significantly steeper than our blue slope, and the overall slope of other quasar templates: $\alpha_\nu = +0.593$ or $\alpha_\lambda = -2.593$. 

The situation for the ultraviolet continuum is very different as the quasar dominates for all luminosities we consider. In the 1215-4000\,\AA\ region, the only significant change with luminosity, and hence redshift, is due to the rest-frame ultraviolet emission line properties; for higher luminosity objects the high-ionization lines become blue-asymmetric and emission-line equivalent widths are smaller (Figs.~\ref{fig:CIV_extrema} and \ref{fig:emlines}). Above redshift $z\approx 2$, the model-flux changes significantly at wavelengths shorter than 1215\,\AA\ as a result of absorption due to the inter-galactic medium.

\section{Summary}

The primary rationale for the information and parametric model presented in this paper is to provide empirically-derived SEDs, covering the rest-frame ultraviolet through near-infrared, for luminous type 1 AGN. 
These SEDs, combined with host galaxy SEDs, can be used to simulate quasar observed-frame optical and infrared colours for various purposes, such as (a) investigating the SEDs of `mean' or typical quasars, (b) generating predictions which incorporate the intra-population properties of quasars/AGN, and (c) estimating the properties of individual objects, including photometric redshifts.

As a function of apparent \textit{i}-band magnitude, there are systematic changes in the average quasar colours at all redshifts, which are well-explained by changes in the strength of the emission lines and the contribution of the host galaxy.
Our model  is capable of reproducing, to within $\sim$0.1 magnitudes, the optical and infrared colours of tens of thousands of quasars across a wide range of redshift ($0<z<5$) and luminosity ($-22 > M_i>-29$).
The variation in the \emph{average} emission line and host galaxy contributions can be well described by simple functions of luminosity which account for the observed changes in colour across $18.1<i_\textrm{AB}<21.5$. To instead model the properties of individual quasars, these parameters can be left free to vary.

Building on work from  spectroscopic studies, the  model encapsulates our current understanding of the intra-population variance in the ultraviolet and optical emission line properties of luminous quasars. This is achieved through just one parameter, which is similar to the ``\ion{C}{IV} distance'' defined by
\citet{2020ApJ...899...96R} and  \citet{2021arXiv210607783R}, and accounts for the observated correlations reported by \citet{Coatman19} and \citet{2021MNRAS.505.3247T}.

The relative flux contribution from stars in the host galaxy to the total quasar SED will always be  larger in the rest-optical than in the rest-ultraviolet. 
Assuming that the quasar optical-continuum slope does not change across the population,
and that the average fractional galaxy contribution has a power law dependence on the total luminosity, we find that bright, $z>2$ quasars from SDSS still have significant (greater than 5 per cent) contributions to their rest-frame optical flux by stellar emission from their host galaxies.
The next decade will bring several new, advanced large-scale surveys of the extragalactic sky. By assuming that the trends observed in the SDSS population continue to fainter luminosities, we make predictions for  the average LSST and \textit{Euclid} colours of soon-to-be-identified  populations of new AGN.

\section*{Acknowledgements}

%The Acknowledgements section is not numbered. Here you can thank helpful
%colleagues, acknowledge funding agencies, telescopes and facilities used etc.
%Try to keep it short.

MJT acknowledges support from CONICYT (Fondo ALMA 31190036), and thanks the IoA, Cambridge, for the award of visitor status while this work was completed.
PCH and MJT acknowledge funding from the Science and Technology Facilities Council via the Institute of Astronomy, Cambridge, Consolidated Grant and a Ph.D. studentship respectively.
MB acknowledges funding from the Royal Society via a University Research Fellowship.
We thank Gordon Richards and an anonymous referee for their careful reading of the manuscript, and Adam Marshall, Matt Auger and Richard McMahon for testing the model code. MJT acknowledges a stimulating examination \textit{viva voce} with Chris Done and Richard McMahon, and thanks Claudio Ricci and Roberto Assef for useful discussions.

The VISTA data flow system, the WFCAM science archive and the VISTA science archive are described in \citet{2004SPIE.5493..411I}, \citet{2008MNRAS.384..637H} and \citet{2012A&A...548A.119C}.
This work  made use of \textsc{astropy} \citep{2013A&A...558A..33A, 2018AJ....156..123A}, \textsc{corner.py} \citep{corner}, \textsc{emcee} \citep{emcee}, \textsc{matplotlib} \citep{Hunter:2007}, \textsc{numpy} \citep{numpy}, \textsc{scipy} \citep{scipy} and \textsc{Q3C} \citep{Q3C}.
This work also made use of the Whole Sky Database (wsdb) created by Sergey Koposov and maintained at the Institute of Astronomy, Cambridge by Sergey Koposov, Vasily Belokurov and Wyn Evans with financial support from STFC and the European Research Council. 
This research has made use of the SVO Filter Profile Service (http://svo2.cab.inta-csic.es/theory/fps/) supported from the Spanish MINECO through grant AYA2017-84089.

This publication makes use of data products from the Wide-field Infrared Survey Explorer, which is a joint project of the University of California, Los Angeles, and the Jet Propulsion Laboratory/California Institute of Technology, and NEOWISE, which is a project of the Jet Propulsion Laboratory/California Institute of Technology. \textit{WISE} and NEOWISE are funded by the National Aeronautics and Space Administration.

Funding for the Sloan Digital Sky Survey IV has been provided by the Alfred P. Sloan Foundation, the U.S. Department of Energy Office of Science, and the Participating Institutions. SDSS-IV acknowledges
support and resources from the Center for High-Performance Computing at
the University of Utah. The SDSS web site is www.sdss.org.

SDSS-IV is managed by the Astrophysical Research Consortium for the 
Participating Institutions of the SDSS Collaboration including the 
Brazilian Participation Group, the Carnegie Institution for Science, 
Carnegie Mellon University, the Chilean Participation Group, the French Participation Group, Harvard-Smithsonian Center for Astrophysics, 
Instituto de Astrof\'isica de Canarias, The Johns Hopkins University, Kavli Institute for the Physics and Mathematics of the Universe (IPMU) / 
University of Tokyo, the Korean Participation Group, Lawrence Berkeley National Laboratory, 
Leibniz Institut f\"ur Astrophysik Potsdam (AIP),  
Max-Planck-Institut f\"ur Astronomie (MPIA Heidelberg), 
Max-Planck-Institut f\"ur Astrophysik (MPA Garching), 
Max-Planck-Institut f\"ur Extraterrestrische Physik (MPE), 
National Astronomical Observatories of China, New Mexico State University, 
New York University, University of Notre Dame, 
Observat\'ario Nacional / MCTI, The Ohio State University, 
Pennsylvania State University, Shanghai Astronomical Observatory, 
United Kingdom Participation Group,
Universidad Nacional Aut\'onoma de M\'exico, University of Arizona, 
University of Colorado Boulder, University of Oxford, University of Portsmouth, 
University of Utah, University of Virginia, University of Washington, University of Wisconsin, 
Vanderbilt University, and Yale University.

\section*{Data availability}

The data underlying this article were accessed from the Sloan Digital Sky Survey,\footnote{\url{https://www.sdss.org/dr16/}}
%the UKIRT Infrared Deep Sky Survey,
the WFCAM science archive,\footnote{\url{http://wsa.roe.ac.uk}}
%the VISTA VIKING Survey,
the VISTA science archive,\footnote{\url{http://horus.roe.ac.uk/vsa/}}
the unWISE catalogue,\footnote{\url{https://catalog.unwise.me}}
and the VHzQ GitHub repo.\footnote{\url{https://github.com/d80b2t/VHzQ/}}
%\footnote{\url{https://github.com/d80b2t/VHzQ/blob/master/data/VHzQs_ZYJHK_WISE_v3.dat}}
The quasar SED model code described in the text is available on GitHub.\footnote{\url{https://github.com/MJTemple/qsogen/}}

%%%%%%%%%%%%%%%%%%%%%%%%%%%%%%%%%%%%%%%%%%%%%%%%%%

%%%%%%%%%%%%%%%%%%%% REFERENCES %%%%%%%%%%%%%%%%%%

% The best way to enter references is to use BibTeX:

\bibliographystyle{mnras}
\bibliography{Thesis_refs, Paper_refs}

%%%%%%%%%%%%%%%%%%%%%%%%%%%%%%%%%%%%%%%%%%%%%%%%%%

%%%%%%%%%%%%%%%%% APPENDICES %%%%%%%%%%%%%%%%%%%%%

\appendix

\section{Note on Temple et al. (2021a)}
\label{sec:HD}

In \citet[][hereafter T21]{2021MNRAS.501.3061T}, we made use of an earlier version of our model SED to investigate the variation in sublimation-temperature dust in the quasar population.
The model employed in T21 made use of a fixed emission-line template, which had been derived from \citet{Francis91}. However, the changes in colour due to the variation in emission-line properties are minimal at the wavelengths $>1$\,\micron\ where we sought to constrain the hot dust emission, and so this has no effect on our results. In particular, the correlation found between the strength of hot dust emission and the blueshift of the \civ\ emission line is not driven by a lack of emission-line variation in the SED model.

Second, the model used in T21 had been calibrated to a bright subset of the fourteenth data release of SDSS, as opposed to the multi--flux-binned DR16 used in this work. This meant that the red power-law slope $\alpha_2$ was -0.16 instead of the value of +0.59 found in this work 
(recall also that we assumed zero host galaxy contribution for the relatively bright objects in T21). This change in slope is reflected in the fact that the median blackbody temperature in T21 was 1280\,K, as opposed to the value of 1240\,K found to be the preferred parameter in this work.
While the power-law slope has a direct impact on the quantitative results of T21 (in particular, the blackbody normalisation \texttt{bbnorm} is re-scaled by a constant factor of around 1.5 to account for the change of blackbody shape), the qualitative results and correlations in T21 remain unchanged. Further details on the robustness of the results of T21 to different model assumptions can be found in the appendices to that work.

\section{Emission line templates}
\label{sec:emlines}

To construct the two emission line templates required in Section~{\ref{sec:lines}}, it is first necessary to identify extreme (in terms of their ultraviolet line properties) objects from among the spectroscopically-observed quasar population.  The SDSS is the only survey that contains sufficiently large numbers of such extreme quasars. However, the SDSS and BOSS spectrographs are  limited in their wavelength range, only taking observed-frame optical data. In order to cover the full $\approx$900\,{\AA} to 3\,{\micron} wavelength range of our SED model, it is therefore necessary to make use of SDSS spectra from sub-populations with differing redshifts, while ensuring that we encapsulate the correlated information between different wavelength regions.
To cover the ultraviolet region at wavelengths $< 4000$\,{\AA}, we therefore make use of SDSS quasars with 
redshifts $z\simeq1.2$ and $\simeq 2.2$ which have been carefully matched using the properties of the emission-line complex that includes {\ion{C}{III}}]\,{$\lambda$}1908. These spectra are then extended to longer wavelengths to include H$\beta$ and {[\ion{O}{III}]}\,$\lambda\lambda$4960,5008.  Finally, H$\alpha$ and the near-infrared lines are included subject to an appropriate scaling to reproduce the observed correlations between ultraviolet and optical line strengths. These  steps are now described in more detail.

\subsection{Rest-frame ultraviolet lines}

We start by defining samples of quasars with spectra from SDSS DR7. 
Specifically, the samples are chosen to display extreme `blueshifts' in the complex of emission lines at \mbox{$\simeq$1908\,\AA}, which is known to correlate with the properties of the {\civ} emission line 
(see appendix A of \citealt{Rankine20}, also fig.~16 of \citealt{Richards11} and sections 5.2 and 6.2 of \citealt{2020MNRAS.496.2565T}).
DR7 spectra are used in preference to BOSS/eBOSS spectra due to the far more accurate spectrophotometry achieved for the DR7 spectra \citep{2013AJ....145...69L, 2016ApJ...831..157M}, meaning the intrinsic large-scale shape of each spectrum is retained.
The DR7 spectra employed are those without evidence of broad absorption lines, with signal-to-noise ratio (S/N) per pixel $\ge 8$ and only a small number of bad pixels. 

Two samples of 200 extreme quasars with redshifts $z\simeq2.2$
were used to generate high S/N composite spectra covering 1050-3000\,{\AA} for each {$\lambda$}1908 morphology.
The effect of the Ly{$\alpha$} forest on quasar spectra at redshift $z\approx2.2$ is small, but, nevertheless, the 1050-1215\,{\AA} region in each composite is corrected for incomplete transmission through the inter-galactic medium using the prescription of \citet{2013MNRAS.430.2067B}. The composite is then extrapolated down to 970\,\AA\, assuming the Ly$\beta$ line is  one-sixth the strength of Ly$\alpha$.

Two  samples of 250 quasars  with redshifts 
 $z\simeq1.2$, with {$\lambda$}1908 morphologies matched to the {$z\simeq2.2$} samples, were then used to generate composite spectra covering 1850-4000\,{\AA}.
 These were co-added over the region of overlap to give an emission line template covering 970-4000\,{\AA} for each {$\lambda$}1908 morphology.
 
\subsection{Rest-frame 4000-5100\,{\AA} region}
 
The wavelength range of the spectrograph used for SDSS DR7 is such that it cannot cover both 1908 and 5008\,{\AA} in the rest-frame of any single object. In order to extend our composite spectra redwards to cover the {H$\beta$} and {[\ion{O}{III}]} lines, we therefore employ a spectral reconstruction technique which uses information from the 2200-4000\,{\AA} region of the spectrum to define the template emission-line spectra over the 4000-5100\,{\AA} region.
 
 A fifth sample of 1666 quasars was defined,  with redshifts $0.73 \le z \le 0.80$ and including the full range of SED-properties.
The same S/N\,$\ge8$, no strong absorption and small number of bad pixel criteria were used as for the construction of the composite spectra in the ultraviolet. 
 Using this sample, a mean-field independent component analysis (MFICA) \citep{Hojen-Sorenson2002Mean-FieldAnalysis, 2013MNRAS.430.3510A}, was performed covering the rest-frame wavelength interval $2200 \le \lambda \le 5100$\,\AA. 
The analysis produced a set of seven MFICA spectral components which can be linearly combined  to reconstruct a spectrum.
The coefficients (`weights') in these linear combinations were left free to vary in order to obtain MFICA reconstructions that reproduced the 2200-4000\,\AA \ spectra of the `extreme' composites generated from the $z \simeq 1.2$ quasar samples.
The 4000-5100\,{\AA} wavelength range of each reconstruction was then associated with the respective composite ultraviolet spectrum.
Confidence in the reliability of the resulting emission-line spectra, covering 970-5100\,\AA, comes from the reproduction of the empirical correlations between the equivalent widths and blueshifts of the \civ\ and [\ion{O}{III}] emission lines found by \citet{Coatman19},
as shown in Fig.~\ref{fig:line_eqws}.

\subsection{H$\alpha$ and near-infrared lines}

Longwards of 5100\,\AA, we include H$\alpha$ and the near-infrared ($\lambda < 2\,\micron$) emission lines using the composite spectrum of \citet{Glikman2006}. These lines are scaled such that the strength of the H$\alpha$ emission line matches the known correlation between the equivalent widths of \civ\ and H$\alpha$, using the catalogue of \citet{Coatman17, Coatman19}.

To remove any remaining high-frequency noise, the composite spectra are smoothed and re-binned to a pixel scale of
\mbox{$\simeq$345}, \mbox{$\simeq$69} and \mbox{$\simeq$490\,\kms} in the ultraviolet, optical and near-infrared regions respectively.
A broken power-law continuum was defined using line-free regions of the composites and subtracted off to give the final form of the emission line templates. This broken power-law is included with the templates as a reference continuum to allow the equivalent widths of the emission lines to be preserved when changing the parametrized continuum in the model.

The end result is a pair of emission-line templates, both covering 970\,\AA\ to 2\,\micron, which span the known diversity of ultraviolet-optical emission-line properties. These are shown in Fig.~\ref{fig:emlines}.

\section{Parameter degeneracies}
\label{sec:corner}

\begin{figure*}
    \centering
    \includegraphics[width=2\columnwidth]{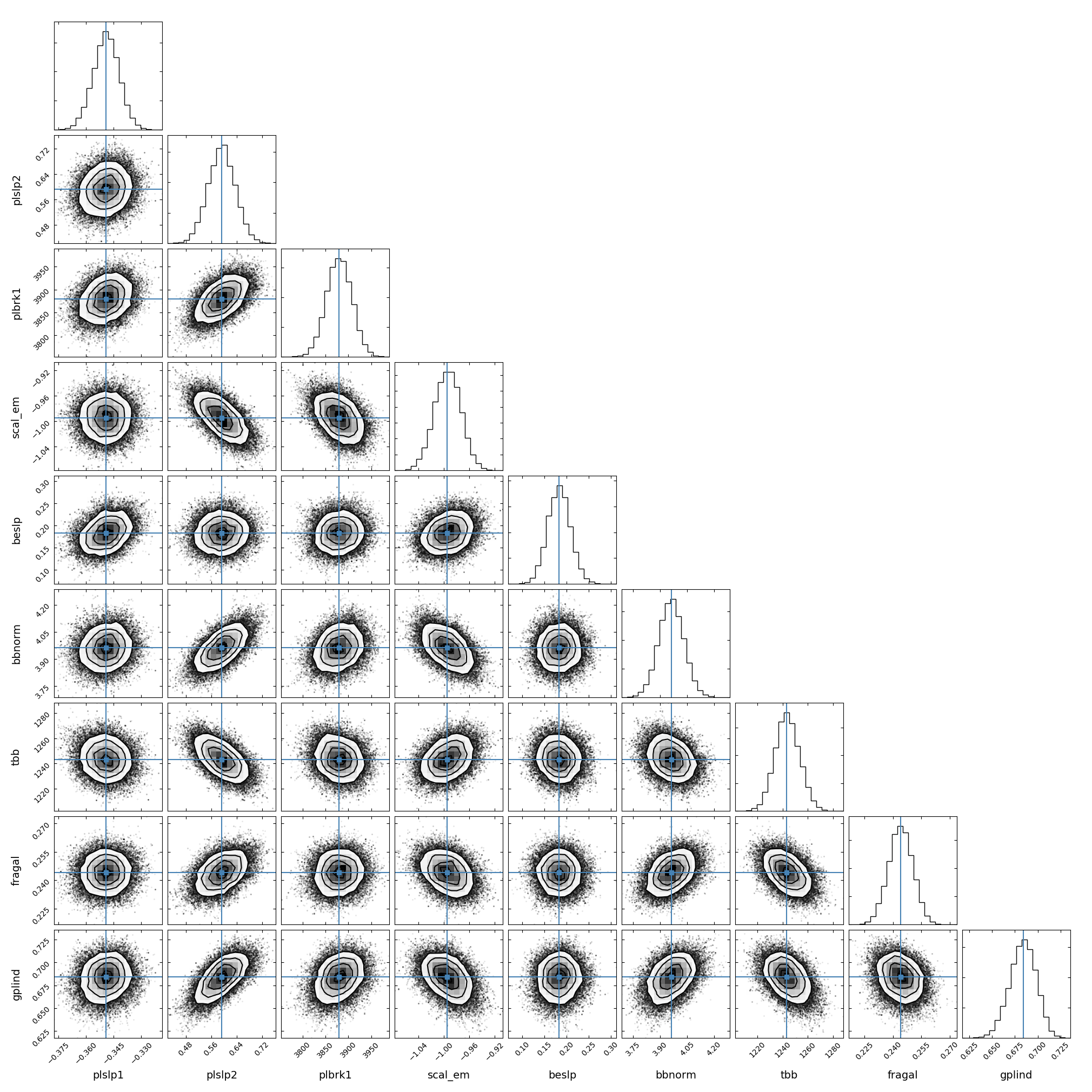}
    \caption{Samples from the posterior distribution of the likelihood surface of SED model parameters defined in equation~\ref{eq:likelihood}. The posterior is seen to be unimodal in all projections. Some  degeneracies can be observed between the strength of luminosity-evolution of the galaxy contribution \texttt{gplind}, the strength of hot dust emission \texttt{bbnorm} and the second power-law slope \texttt{plslp2}, but otherwise all parameters are well-determined. The maximum-likelihood solution is shown in blue.}
    \label{fig:MCMC_corner}
\end{figure*}

The posterior distribution of the likelihood surface of model parameters is shown in Fig.~\ref{fig:MCMC_corner}.
Degeneracies are to be expected between certain parameters:
for example the strength of luminosity-evolution of the galaxy contribution \texttt{gplind}, the strength of hot dust emission \texttt{bbnorm} and the second power-law slope $\alpha_2$ can all effect the overall shape of the rest-frame near-infrared SED.

%%%%%%%%%%%%%%%%%%%%%%%%%%%%%%%%%%%%%%%%%%%%%%%%%%

% Don't change these lines
\bsp    % typesetting comment
\label{lastpage}
\end{document}

%% file: thesis_macros.tex
\providecommand{\ion}[2]{\mbox{\textrm{#1\,\textsc{#2}}}}

\def\km{\textrm{\thinspace km}}

\def\micron{\hbox{$\mu$m}}
\def\Mpc{\textrm{\thinspace Mpc}}

\def\ps{\textrm{\thinspace s^{-1}}}

\def\s{\textrm{\thinspace s}}

\def\kms{\mbox{$\km\ps$}}
\def\kmpspMpc{\mbox{$\km\ps\Mpc^{-1}$}}

\def\ps{\mbox{$\s^{-1}\,$}}

\def\lya{\mbox{\textrm{Ly}$\alpha$}}
\def\lyb{\mbox{\textrm{Ly}$\beta$}}
\def\lyg{\mbox{\textrm{Ly}$\gamma$}}

\def\civ{\mbox{\textrm{C\,\textsc{iv}}}}

\def\h0{\mbox{\textrm{H}$^0$}}
\DeclareMathAlphabet{\vib}{OML}{cmm}{m}{it}